\documentclass[twocolumn,aps,superscriptaddress,preprintnumbers]{revtex4}

\usepackage{graphicx} 

\usepackage{braket}

\usepackage{bm} 

\usepackage{hyperref}
\hypersetup{    colorlinks=true,       
    linkcolor=blue,        
    citecolor=blue,        
    filecolor=blue,        
    urlcolor=blue,         
    runcolor=blue
}

\usepackage{xcolor}

\begin{document}

\title{Theory of Nanoscale Organic Cavities:\\ The Essential Role of Vibration-Photon Dressed States}

\author{Felipe Herrera}
\email{felipe.herrera.u@usach.cl}
\affiliation{Department of Physics, Universidad de Santiago de Chile, Av. Ecuador 3493, Santiago, Chile}

\author{Frank C. Spano}
\email{spano@temple.edu}
\affiliation{Department of Chemistry, Temple University, Philadelphia, Pennsylvania 19122, USA}


\begin{abstract}
The interaction of organic molecules and molecular aggregates with electromagnetic fields that are strongly confined inside optical cavities within nanoscale volumes, has allowed the observation of exotic quantum regimes of light-matter interaction at room temperature, for a wide variety of cavity materials and geometries. Understanding the universal features of such organic cavities represents a significant challenge for theoretical modelling, as experiments show that these systems are characterized by an intricate  competition between coherent and dissipative processes involving entangled nuclear, electronic and photonic degrees of freedom. In this review, we discuss a new theoretical framework that can successfully describe organic cavities under strong light-matter coupling. The theory combines standard concepts in chemical physics and quantum optics to provide a microscopic description of vibronic organic polaritons that is fully consistent with available experiments, and yet is profoundly different from the common view of organic polaritons. We show that by introducing a new class of vibronic polariton wave functions with a photonic component that is dressed by intramolecular vibrations, the new theory can offer a consistent solution to some of the long-standing puzzles in the interpretation of organic cavity photoluminescence. Throughout this review, we confront the predictions of the model with spectroscopic observations, and describe the conditions under which the theory reduces to previous approaches.  We finally discuss possible extensions of the theory to account for realistic complexities of  organic cavities such spatial inhomogeneities and the multi-mode nature of confined electromagnetic fields. 
\end{abstract}

\maketitle

 Since the early demonstrations of strong light-matter coupling and polariton formation in organic microcavities over two decades ago  \cite{Lidzey1998,Lidzey1999,Lidzey2000}, the optical properties of organic polaritons have been intensely studied in a variety of nanoscale optical cavities with different geometries and material composition  \cite{Hobson2002,Tischler2005,Holmes2007,kena-cohen2008,Kena-Cohen2010,Hutchison2013,Bellessa2014,Schartz2011,Kena-Cohen2013,Mazzeo2014,Cacciola2014,Gambino2015,George2015-farad,Long2015,Muallem2015,George2015,Shalabney2015,Saurabh2016,Chikkaraddy:2016aa}. Despite fundamental questions regarding the microscopic behaviour of organic polaritons that are still open, pioneering applications of these systems for the control of chemical reactions  \cite{Hutchison:2012,Simpkins2015,Herrera2016,Galego2016}, as well as enhancement of transport  \cite{Andrew2000,Hutchison2013,Feist2015,Schachenmayer2015,Orgiu2015,yuen2016} and nonlinear optical properties \cite{Herrera2014,bennett2016novel,Kowalewski2016,kowalewski2016cavity} of organic semiconductors are paving the way for the development of optoelectronic devices that can be enhanced by quantum optics. 
 
The study of organic cavities originally started as a promising variation of inorganic semiconductor microcavities  \cite{Agranovich1997}. It is therefore useful to first discuss some of the relevant differences and similarities between organic polaritons and their inorganic counterparts. Inorganic semiconductor microcavities require high-quality dielectric mirrors ($Q> 10^5$), highly ordered samples, and cryogenic temperatures to reduce the loss of polariton coherence induced by photon loss and exciton scattering  \cite{Skolnick1998,Vahala2003}. The linear optical response of inorganic microcavities is largely dominated by the so-called polariton splitting in the transmission, reflection and absorption spectra, which is centered at the bare exciton frequency for resonant light-matter coupling. In perfectly ordered inorganic samples, there are no light-matter states in the frequency region between the lower and upper polariton peaks. However,  lattice disorder in realistic samples can lead to the formation of localized polariton states in the vicinity of the bare excitonic resonance  \cite{Muller2000}, which can also absorb and emit light. The relaxation of inorganic polariton relaxation is dominated by non-radiative decay via perturbative polariton-phonon scattering  \cite{Skolnick1998, Deng2010}, and also radiative relaxation due to the leakage of the cavity photon that forms part of the polariton wave function, through the cavity mirrors into the far field. Photons that leak out of the cavity can be detected in reflection, transmission and photoluminescence experiments  \cite{Savona1996}, providing information about the photonic component of the parent polariton state from which they  originated  \cite{Savona1999}. For a system with small structural disorder, the overall polariton relaxation dynamics is thus determined by a competition of timescales between phonon scattering and photon leakage  \cite{Muller2000,Tassone1997}. At higher excitation densities, polariton-polariton scattering becomes important as well  \cite{Deng2010}. 

Organic microcavities differ from inorganic systems in several practical and fundamental ways. For example, with organic cavities it is not necessary to cool the material to very low temperatures in order to reach the so-called strong coupling regime of cavity quantum electrodynamics \cite{Mabuchi2002}, as is the case for inorganic samples. Organic polaritons can be prepared at room temperature. Another important difference is the much shorter lifetime of cavity photons in typical organic cavities with metallic mirrors in comparison with inorganic systems having dielectric mirrors. Metallic microcavities and plasmonic nanocavities can have quality factors as low as $Q\sim 10$  \cite{Hobson2002,George2015-farad,Chikkaraddy:2016aa}, so that photons can quickly escape the cavity volume into the far field in sub-picosecond timescales \cite{Hobson2002,George2015-farad}, which is shorter than the timescales for intramolecular vibrational relaxation \cite{May-Kuhn}, and much shorter than the photon lifetime in inorganic microcavities \cite{Muller2000}. In such lossy cavities, polariton formation is only supported transiently by a very strong coherent light-matter coupling. In organic cavities,  the so-called vacuum Rabi frequency \cite{Tischler2007}, which quantifies the strength of light-matter coupling, can reach record values in the range  $0.1-1$ eV$/\hbar$ \cite{Schartz2011,Kena-Cohen2013,Cacciola2014,Gambino2015,Mazzeo2014,Ebbesen2016},  several orders of magnitude larger than the values achieved with inorganic microcavities \cite{Skolnick1998,Vahala2003}, atomic cavities \cite{Mabuchi2002}, or superconducting resonators \cite{Blais2004,Wallraff2004}. 

In addition to the difference in relaxation timescales and light-matter coupling strength, another feature that distinguishes organic cavities from their inorganic analogues is the strength of electron-phonon coupling. It is well-known that in addition to low-frequency environmental phonons  \cite{May-Kuhn}, the transport  \cite{Coropceanu2007} and photophysical properties  \cite{Spano2010} of organic semiconductors are heavily influenced by strong vibronic coupling between electrons and high-frequency intramolecular vibrations. For many chromophores, vibronic coupling involving high-frequency modes can even compete in strength with the coherent Rabi coupling inside a nanoscale cavity. This is strikingly different from inorganic microcavities, where electron-vibration coupling is only perturbative  \cite{Skolnick1998,Vahala2003}. The strong electron-vibration coupling in organic materials gives a vibronic structure to the spectral signals of organic polaritons that has been well established experimentally  \cite{Holmes2004,Holmes2007,Coles2011,Virgili2011,George2015-farad,Ebbesen2016}.  

\begin{figure}[t]
\includegraphics[width=0.5\textwidth]{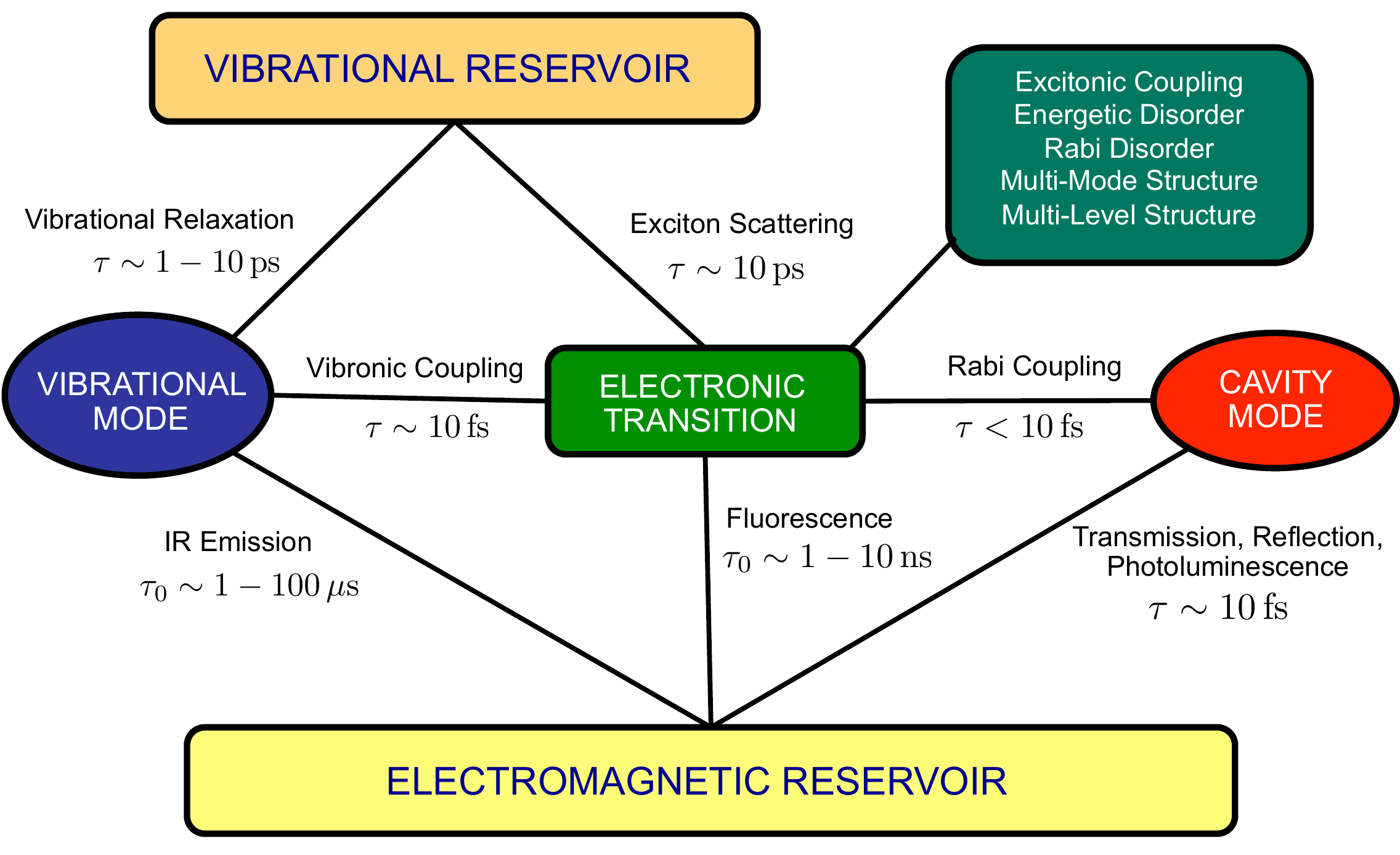}
\caption{Illustration of the relevant degrees of freedom involved in the description of organic cavities and their characteristic dynamical timescales. $\tau$ denotes an intracavity magnitude and $\tau_0$ denotes a free-space value.}
\label{fig:dof}
\end{figure}

In order to model organic cavities, we therefore need a theory that is able to describe the competition between the multiple coherent and dissipative processes that determine the observable properties of organic polaritons, such their absorption and emission spectra. We have  recently developed a theoretical framework \cite{Spano2015,Herrera2016,Herrera2017-PRA,Herrera2017-PRL} that combines elements from condensed matter theory and quantum optics, to describe the interaction between nuclear, electronic and photonic degrees of freedom illustrated in Fig. \ref{fig:dof}. The theory builds on the recently introduced Holstein-Tavis-Cummings (HTC) model \cite{Cwik2014}, together with a Markovian treatment of dissipation. 
The theoretical framework based on the HTC model reduces to previous quasiparticle theories of organic polaritons  \cite{Agranovich2003,Litinskaya2004,Litinskaya2006,Cwik2016,Fontanesi2009,Mazza2009} under certain conditions. However, the theory is more general, as it allows us to provide a consistent interpretation for some of the {\it conundrums} \cite{George2015-farad} observed in the absorption and emission spectra of organic cavities \cite{Hobson2002,Ebbesen2016}. 

Throughout this review, we show that the HTC model offers a possible solution to the conundrums by introducing a new type of hybrid light-matter state, the {\it dark vibronic polariton} \cite{Herrera2017-PRL}, a polariton eigenstate that cannot be accessed spectroscopically by absorption from the absolute ground state of the cavity, but yet can strongly emit light into the far field by direct photon leakage. The emitted photons can have {\it lower energy} than the dark vibronic polariton state from which they originated, something that is not possible within previous organic polariton theory \cite{Agranovich2003,Litinskaya2004,Litinskaya2006,Cwik2016,Fontanesi2009,Mazza2009}. We show that this profoundly different view of vibronic polariton emission is valid for any value of the Rabi coupling strength, which may have important implications for macroscopic coherence phenomena such as condensation \cite{Deng2010,Daskalakis2017} and lasing \cite{Kena-Cohen2010,Ramezani2017}.

\section{Holstein-Tavis-Cummings Model}

\begin{figure}[t]
\includegraphics[width=0.45\textwidth]{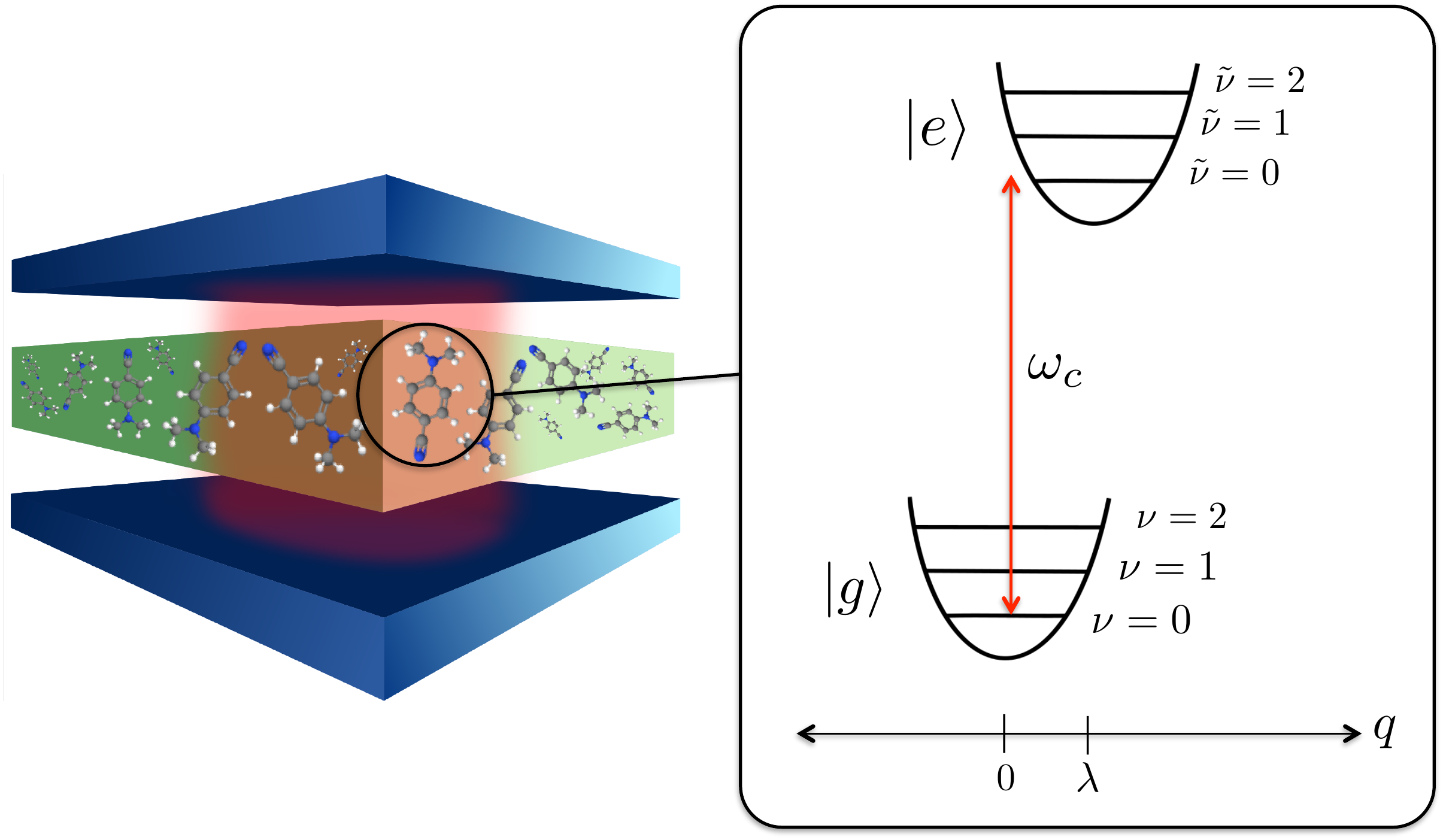}
\caption{Illustration of a planar organic microcavity with an ensemble of organic emitters, each being represented by displaced harmonic oscillator potentials in a dimensionless nuclear coordinate $q$ associated with a high-frequency intramolecular vibrational mode. As a result of vibronic coupling, the nuclear potential minimum in the excited electronic state $\ket{e}$ is displaced by $\lambda$ from the equilibrium configuration in the ground state $\ket{g}$. Transitions between vibrational eigenstates in different potentials are driven by a cavity field with a mode frequency $\omega_c$}
\label{fig:displaced oscillator}
\end{figure}

We begin our discussion by introducing the Holstein-Tavis-Cummings (HTC) model, which describes an ensemble of $N$ organic emitters interacting with a single cavity mode. The HTC Hamiltonian can be written as \cite{Cwik2014,Spano2015,Herrera2016}
\begin{eqnarray}\label{eq:HTC}
\hat{\mathcal{H}} &=& \omega_c \,\hat a^\dagger \hat a+ \omega_{\rm v}\sum_{n=1}^N\hat b_n^\dagger \hat b_n\nonumber\\
&&+\sum_{n=1}^N\left[\omega_e+\omega_{\rm v}\lambda(\hat b_n+\hat b_n^\dagger)\right]\ket{e_n}\bra{e_n},\nonumber\\
&&+\frac{\Omega}{2}\sum_{n=1}^N(\ket{g_n}\bra{e_n}\hat a^\dagger +\ket{e_n}\bra{g_n}\hat a),
\end{eqnarray}
where $\omega_e =\omega_{00}+ \omega_{\rm v}\lambda^2$ is the vertical Frank-Condon transition frequency, with $\omega_{00}$ being the frequency of the zero-phonon (0-0) vibronic transition, $\omega_{\rm v}$ is the intramolecular vibrational frequency and $\lambda^2$ is the Huang-Rhys factor  \cite{Spano2010}, which quantifies the strength of vibronic coupling within the displaced oscillator model illustrated in Fig. \ref{fig:displaced oscillator}. The operator $\hat b_n$ annihilates one quantum of vibration on the $n$-th emitter. The operator $\hat a$  annihilates a cavity photon with frequency $\omega_c$, and $\Omega$ is the vacuum Rabi frequency for a single molecular emitter, which  can reach several tens of meV in nanoscale plasmonic systems  \cite{Chikkaraddy:2016aa}.

The HTC model in Eq. (\ref{eq:HTC}) describes either individual molecules or molecular aggregates interacting with a confined electromagnetic field for which dispersion can be neglected within the photon linewidth. We assume that all molecules (or aggregates) are {\it equally} coupled to the electric field of a single broadened quasi-mode of the cavity $\omega_c(k)$, where $k$ is a mode parameter that determines the photon energy, such as the in-plane wave vector for planar mirrors. Since metallic cavities are highly dispersive and the electric field is not spatially homogeneous throughout the organic material, the HTC model in its current form can only provide a qualitative description of the dispersive behaviour of organic polaritons. However, we expect the predictions of the HTC model to become accurate for  systems where the electronic transition frequency of the molecular system is resonant with the region of the cavity dispersion diagram with the largest density of states, which is less dispersive. For planar microcavities with dispersion of the form \cite{Savona1999} $\omega_c(k)=\omega_c\sqrt{1+(k/k_0)^2}$, the region with large mode density occurs near $k=0$, i.e., at normal incidence. 

Real organic samples are highly disordered, and energetic disorder in the electronic transition frequency is well-known to influence the photophysics of organic semiconductors  \cite{Fidder1991a,Spano2010}. In organic cavities, there is also a large degree of structural disorder associated with the orientation of the molecular transition dipoles relative to the spatially inhomogeneous cavity field, which leads to the coexistence of molecular dipoles that remain in the weak coupling regime due to inefficient light-matter coupling, and molecular dipoles that can strongly couple to the cavity field and form polariton states   \cite{Agranovich2003,Gonzalez2013}. We have recently shown that the qualitative predictions of the homogeneous HTC model in Eq. (\ref{eq:HTC}) can hold in systems where the light-matter coupling strength  exceeds the typical energy disorder widths  \cite{Spano2015,Herrera2016,Herrera2017-PRA,Herrera2017-PRL}, which is the case for several experimental realizations of organic cavities    \cite{Hobson2002,Coles2011,Virgili2011,Schwartz2013}. 

\subsection{The Single-Particle Approach}

\begin{figure}[t]
\includegraphics[width=0.5\textwidth]{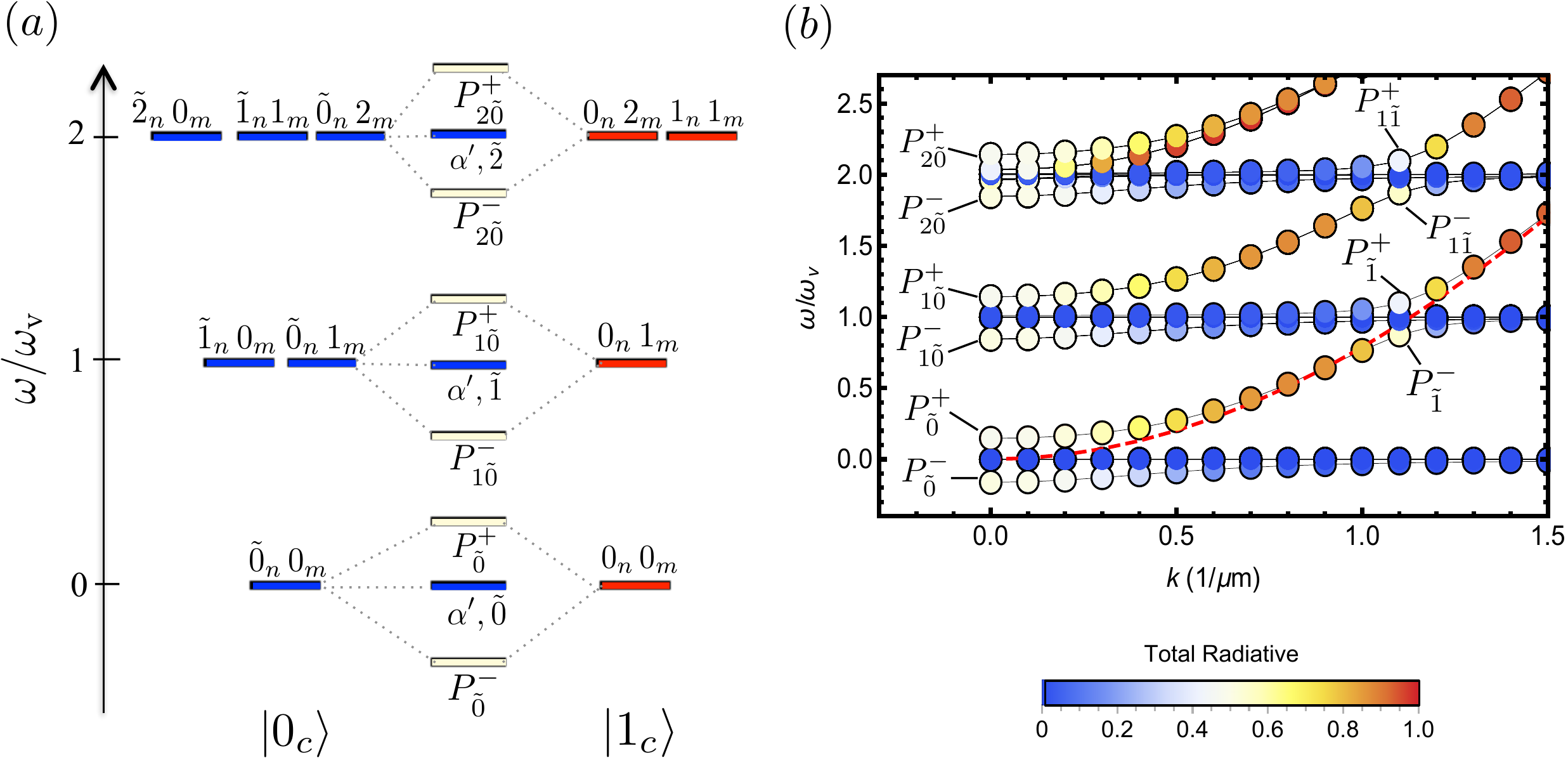}
\caption{Spectrum of the Holstein-Tavis-Cummings model for small Rabi couplings $\sqrt{N}\Omega/\omega_{\rm v}\ll 1$. (a) Energy level diagram of the single-excitation manifold in the small Rabi limit $\sqrt{N}\Omega/\omega_{\rm v}\ll 1$, displaying the dominant Rabi couplings between material states with $n_c=0$ cavity photons and dressed states with $n_c=1$ photon. $\ket{ G}$ is the absolute ground state of the system. The notation $\tilde\nu_n\nu_m$ indicates the vibronic and vibrational configuration of molecules $n$ and $m$ in the ensemble, and $\nu_n\nu_m$ the vibrational configuration of photon dressed states. States $(\alpha' \tilde \nu)$ corresponds to collective dark vibronic states that are non-symmetric with respect to particle permutation. Single-particle $P_{\tilde \nu}^\pm$ and two-particle polaritons $P_{\nu \tilde \nu}^\pm$ are shown.   (b) Polariton dispersion curves for $N=10$, $\sqrt{N}\Omega/\omega_{\rm v}=0.4$ and $\lambda^2=0.5$. Symbols are colored according to the total radiative component of the corresponding polariton eigenstate. The red-dashed line corresponds to the empty-cavity dispersion. We use the resonance condition $\omega_{c}=\omega_{00}$ at normal incidence. Frequency is shown relative to $\omega_c$.  }
\label{fig:HTC spectrum}
\end{figure}

In Fig. \ref{fig:HTC spectrum}(a), we illustrate the spectrum of the HTC Hamiltonian $\hat{\mathcal{H}}$ for an ensemble of $N$ molecular emitters in a cavity with Rabi coupling parameter $\sqrt{N}\Omega/\omega_{\rm v}\ll 1$. We show only those states within the manifold that involves up to one vibronic excitation or one cavity photon, with any number of vibrational excitations. This is the analogue of the one-excitation manifold of cavity QED in quantum optics \cite{Carmichael-book2}. The absolute ground state of the cavity is given by the product form 
\begin{equation}\label{eq:ground state}
\ket{G}\equiv\ket{g_10_1, g_20_2,\ldots,g_n0_n,\ldots g_N0_N; 0_c},
\end{equation}
which describes a cavity without electronic, vibrational or photonic excitations. The state $\ket{g_n0_n}$ describes the $n$-th molecular emitter in the ground vibrational state ($\nu=0$) of its ground electronic potential, and the photon vacuum is denoted as $\ket{0_c}$.

Given that within the HTC model the confined electromagnetic field is unable to distinguish between individual molecular emitters, a single photon state can couple to the {\it delocalized} vibronic excitation 
\begin{equation}\label{eq:SP material}
\ket{\alpha_0,\tilde \nu,0_c}\equiv\left(\frac{1}{\sqrt{N}}\sum_{n=1}^N\ket{g_10_1,\ldots e_n\tilde \nu_n, \ldots g_N 0_N}\right)\otimes \ket{0_c},
\end{equation}
where $\alpha_0$ is a permutation quantum number that describes the totally-symmetric character of the superposition with respect to particle permutation. We use the displaced oscillator notation $\tilde \nu_n \geq 0$ from Fig. \ref{fig:displaced oscillator} to describe the vibrational quantum number of the $n$-th molecule in its excited electronic potential. 
 In the language of molecular aggregates, state $\ket{\alpha_0,\tilde \nu,0_c}$ would correspond to the {\it bright exciton} \cite{Saikin2013a} state. We note however, that in contrast to the Frenkel exciton formalism \cite{Spano2010,Saikin2013a}, permutation symmetry within the homogeneous HTC model emerges naturally from the inability of the electric field to distinguish between individual emitters, not as the result of intermolecular coupling in systems with translational invariance, such as molecular aggregates or molecular crystals. 

For the small value of the Rabi coupling parameter $\sqrt{N}\Omega/\omega_{\rm v}\ll 1$ considered in Fig. \ref{fig:HTC spectrum}, the resonant coupling between the photon state $\hat a^\dagger \ket{G}\equiv \ket{g_10_1,g_20_2\ldots,g_N0_N;1_c}$ and the material state $\ket{\alpha_0,\tilde \nu,0_c}$ leads to the formation of {\it diabatic} single-particle polariton states given by
\begin{equation}\label{eq:SP polariton}
\ket{P_{\tilde \nu}^\pm} = \frac{1}{\sqrt{2}}\left(\ket{\alpha_0,\tilde \nu,0_c}\pm\ket{g_10_1,g_20_2\ldots,g_N0_N;1_c}\right), 
\end{equation}
which have diabatic energies $
E_{\tilde \nu}^\pm  = \omega_{00}+\tilde \nu \omega_{\rm v}\pm \sqrt{N}	\,\Omega|\langle 0|\tilde \nu\rangle |/2$ when $\omega_{c}= \omega_{00}+\tilde \nu\omega_{\rm v}$. These diabatic states define a conventional polariton doublet split by $\Delta E_{\tilde \nu} = \sqrt{N}\,\Omega|\langle 0|\tilde \nu\rangle |$.
The dispersion diagram in Fig. \ref{fig:HTC spectrum}(b) shows that at normal incidence ($k=0$) the resonance condition $\omega_c=\omega_{00}$ leads to the formation of a conventional single-particle polariton splitting $\Delta E_{\tilde 0}$. For higher wave vectors, there can be additional single-particle anti-crossings when the cavity obeys the resonance condition $\omega_c=\omega_{00}+\bar \nu \omega_{\rm v}$. For instance, in Fig. \ref{fig:HTC spectrum}(b), there is a splitting $\Delta E_{\tilde 1}$ when the in-plane wave vector is $k\approx 1.1 $ $\mu$m$^{-1}$.
 
Since the same single-photon state $\hat a^\dagger \ket{G}$ can simultaneously couple to two or more vibronic excitations having different number of displaced vibrational quanta  $\tilde \nu$ (sidebands), the most general diabatic single-particle eigenstate of the HTC model can be written as
\begin{equation}\label{eq:general SP}
\ket{P_j} = \left(\sum_{\tilde \nu}C_{j \tilde \nu}\ket{\alpha_0,\tilde \nu,0_c}\right)+D_j \ket{g_10_1,g_20_2\ldots,g_N0_N;1_c}, 
\end{equation}
where the values of the orthonormal vibronic and photonic coefficients $C_{j \tilde \nu}$ and $D_j$ are obtained from diagonalization of the HTC Hamiltonian in the single-particle basis.  Several experimental works have discussed the vibronic structure of organic polariton spectra on the basis of such single-particle diagonalizations  \cite{Holmes2004,Holmes2007,Coles2011,Virgili2011,Fontanesi2009,Mazza2009}. The defining feature of this single-particle theoretical approach to describe the spectroscopy of organic cavities is the definition of a unique single-photon state in the problem, in which all the molecules in the ensemble are in the vibrationless ground state $(\hat a^\dagger \ket{G})$. This restriction has immediate consequences in the predicted polariton emission dynamics. For example, when the photonic component of the state $\ket{P_j}$ in Eq. (\ref{eq:general SP}) is emitted into the far field by leakage out of the strong light-matter interaction volume, the emitted photon can be detected at the {\it same energy} of the parent polariton state, because the organic material is projected into the vibrationless ground state $\ket{g_10_1,g_20_2\ldots,g_N0_N}\otimes \ket{0_c}$.  We show in the next sections that going beyond the single-particle description of organic polaritons results significantly changes this description of radiative polariton decay.

Also within the single-particle approximation, the HTC model predicts the existence of $N-1$ {\it collective} vibronic states \cite{Herrera2016,Gonzalez2016} that remain uncoupled to the cavity field because they are non-symmetric with respect to particle permutations. These so-called dark collective states can be written as
\begin{equation}\label{eq:dark exciton}
\ket{\alpha',\tilde \nu,0_c}\equiv\sum_{n=1}^N c_{\alpha' n}\ket{g_10_1,\ldots e_n\tilde \nu_n, \ldots g_N 0_N}\otimes \ket{0_c},
\end{equation}
where $c_{\alpha'n }$ are orthonormal coefficients that depend on the permutation symmetry quantum number $\alpha'\neq \alpha_0$, and satisfy the condition $\sum_nc_{\alpha'n}=0$. 
In the language of molecular aggregates, state $\ket{\alpha\neq \alpha_0,\tilde \nu,0_c}$ would correspond to a {\it dark exciton} state \cite{Saikin2013a}. Since they do not participate in the light-matter coupling within this single-particle approximation, they have the same energy as the bare vibronic transition, as Fig. \ref{fig:HTC spectrum}(a) illustrates. Despite being uncoupled from the cavity field, these dark states are nevertheless collective, a feature that is well-known from the Tavis-Cummings model in quantum optics  \cite{Garraway2011}. The Tavis-Cummings model of optical cavities also predicts that system inhomogeneities such energetic or Rabi disorder break the permutation symmetry of the problem, which results in the mixing of uncoupled dark states with polariton eigenstates \cite{Lopez2007}. We therefore expect a similar phenomenology to occur in disordered organic cavities within the HTC model \cite{Herrera2016}. 

\subsection{Beyond Single-Particle Theories}
In order to describe the eigenstates of the HTC model in Eq. (\ref{eq:HTC}) beyond the single-particle approximation, we need to briefly revisit the role of permutation symmetry in the classification of material and photonic states in a homogeneous system. We note that the HTC Hamiltonian can be written in the form $\hat \mathcal{H}=\omega_c\hat a^\dagger\hat a + \sum_n \hat H_n$, where $\hat H_n$ corresponds to the electron-vibration and electron-photon coupling terms. For a homogeneous molecular ensemble, i.e., no energetic or Rabi disorder, we have that $\hat H_n = \hat H_m$ for any $n\neq m$, which makes  $\hat \mathcal{H}$ invariant under permutation of emitters. The eigenstates of the HTC model in Eq. (\ref{eq:HTC}) therefore must have a well-defined particle permutation symmetry. 

It is nevertheless convenient to treat the permutation symmetry of the vibrational degrees of freedom of the ensemble separately from the permutation symmetry of the electronic degrees of freedom, preserving the overall symmetry of HTC polariton eigenstates. For example, we can introduce a purely vibrational collective excitation (phonon) in the ground manifold of the form
\begin{equation}\label{eq:phonon wave}
\ket{ \beta,\nu,0_c} =\left(\sum_{n=1}^Nc_{\beta n}\ket{g_10_1\ldots,g_n\nu_n,\ldots,g_N0_N}\right)\otimes\ket{0_c},
\end{equation}
where $\nu_n\geq 1$ and $\beta$ is the permutation quantum numbers associated with the nuclear degrees of freedom of the ensemble. There are $N$ possible values for $\beta$, of which  $\beta=\alpha_0$ describes the totally-symmetric superposition, and the remaining $N-1$ values $\beta'\neq \alpha_0$ are associated with non-symmetric combinations. In the language of exciton theory, the collective vibrational excitation in Eq. (\ref{eq:phonon wave}) would correspond to an intermolecular phonon \cite{Grover1970,Spano2010}. However, we stress that the collective nature of vibrational excitations in a cavity does not emerge from the long-range electrostatic interaction between emitters as is the case in molecular aggregates \cite{May-Kuhn}, but from the indistinguishability of the molecular emitters coupled to the same electric field. 

\subsection{Two-Particle Polaritons}

The collective phonon state $\ket{ \beta,\nu,0_c}$ in Eq. (\ref{eq:phonon wave}) belongs to the ground state manifold of the HTC model. In the first excitation manifold (see Fig. \ref{fig:HTC spectrum}), we can define the phonon-photon dressed state 
\begin{equation}\label{eq:photon-phonon}
\ket{\beta,\nu,1_c} = \hat a^\dagger\ket{\beta,\nu,0_c},
\end{equation}
having diabatic energy $\omega_c +\nu \omega_{\rm v}$, with $\nu\geq 1$. Including vibrationally-excited single photon dressed states into the description of organic cavities significantly departs from the single-particle approach to describe vibronic polaritons \cite{Agranovich2003,Agranovich2005,Litinskaya2004,Litinskaya2006,Fontanesi2009,Mazza2009}. The metastable state $\ket{\beta,\nu,1_c}$ can decay both radiatively via photon leakage and non-radiatively via vibrational relaxation. However, as we discuss later in more detail, for systems where vibrational relaxation is slower than radiative relaxation, we expect that vibrationally-excited dressed photon states in Eq. (\ref{eq:photon-phonon}) can significantly contribute to the light-matter dynamics.

The phonon-dressed Fock state $\ket{\beta,\nu,1_c}$ in Eq. (\ref{eq:photon-phonon}) can exchange energy resonantly with the two-particle diabatic material state \cite{Spano2010} 
\begin{eqnarray}\label{eq:TP state}
\lefteqn{\ket{\alpha_0\beta,\tilde \nu' \nu,0_c} =}\\
&&  \sum_{n\neq m, m}^N\frac{c_{\beta m}}{\sqrt{N-1}}\ket{g_10_1,\ldots,e_n\tilde \nu'_n, g_m\nu_m,\ldots,g_N0_N}\otimes\ket{0_c},\nonumber
\end{eqnarray}
which describes an ensemble where the $n$-th molecule is electronically excited in the state $\ket{e_n\tilde \nu_n'}$ and the $m$-th molecule is  vibrationally excited in state $\ket{g_m\nu_m}$, with $\nu_m\geq 1$, while the remaining molecular emitters remain in their absolute ground state. For each of the $N$ ways to place a single vibrational excitation $\nu_m$ among the molecules in the ensemble, there are $N-1$ distinct ways of placing the vibronic excitation $\tilde\nu'_n$. Within the homogeneous HTC model in Eq. (\ref{eq:HTC}), these different possibilities are degenerate and have diabatic energy $\omega_{00}+(\tilde\nu'+\nu)\omega_{\rm v}$. Given this degeneracy, we are free to choose the coefficients of the superposition in Eq. (\ref{eq:TP state}), within the constraints imposed by the fermionic character of the excitations on the same molecular emitter \cite{Herrera2017-PRA}. We  focus on the vibronic superposition that is totally-symmetric with respect to particle permutation, setting equal amplitudes $1/\sqrt{N-1}$ \cite{Herrera2017-PRA}, and choose vibrational superposition coefficients $c_{\beta m}$ to be determined the same permutation quantum number $\beta$ as the phonon-photon dressed state $\ket{ \beta,\nu,0_c}$ in Eq. (\ref{eq:phonon wave}). 

We show in Figure \ref{fig:HTC spectrum}, that the HTC model allows the resonant coupling of two-particle material state $\ket{\alpha_0\beta,\tilde \nu' \nu,0_c} $ with phonon-dressed states $\ket{\beta,\nu,1_c}$, preserving the  permutation quantum number $\beta$, to form diabatic two-particle polariton states of the form \cite{Herrera2017-PRA,Herrera2017-PRL}
\begin{equation}\label{eq:TP polariton}
\ket{P_{ \nu \tilde\nu'}^\pm,\beta} = \frac{1}{\sqrt{2}}\left(\ket{\alpha_0\beta,\tilde\nu'\nu,0_c}\pm \ket{\beta,\nu,1_c}\right),
\end{equation}
having diabatic energies $ E_{\nu\tilde \nu'}^{\pm} = \omega_{00}+(\tilde \nu'+\nu)\omega_{\rm v}\pm \sqrt{(N-1)}\,\Omega\,|\langle 0|\tilde \nu' \rangle|/2$, when the resonance condition $\omega_{c}=\omega_{00} + \tilde \nu'\omega_{\rm v}$ holds\footnote{The resonance condition $\omega_{c}=\omega_{00} + \tilde \nu'\omega_{\rm v}$ replaces the expression $\omega_{c}=\omega_{00} $ in Refs.  \cite{Herrera2017-PRA,Herrera2017-PRL}}. For example, in Fig. \ref{fig:HTC spectrum}(b) we show the two-particle polariton splittings $\Delta E_{1\tilde 0}$ and $\Delta E_{2\tilde 0}$, which form when $\omega_c=\omega_{00}$, and the two-particle splitting $\Delta E_{1\tilde 1}$ forming when $\omega_c=\omega_{00}+\omega_{\rm v}$. The possibility of forming these two-particle polaritons in organic cavities is not captured by commonly used single-particle theories \cite{Agranovich2003,Agranovich2005,Litinskaya2004,Litinskaya2006,Fontanesi2009,Mazza2009,Michetti2008,Cwik2014}.

The two-particle polariton state $\ket{P_{ \nu \tilde\nu'}^\pm,\beta}$ in Eq. (\ref{eq:TP polariton}) does not have a transition dipole moment that connects with the absolute ground state of the cavity, i.e.,  $\braket{G|\hat \mu|P_{ \nu \tilde\nu'}^\pm,\beta}=0$, with $\hat \mu$ being the electric dipole operator. Following the definition introduced in previous work \cite{Herrera2017-PRA,Herrera2017-PRA}, we can then refer to two-particle polariton states $\ket{P_{ \nu \tilde\nu'}^\pm,\beta}$ as a type of {\it dark vibronic polaritons} that exists in the system for relatively small values of the Rabi coupling parameter $\sqrt{N}\Omega/\omega_{\rm v}\ll 1$. We later show that for different strengths of the Rabi coupling, other types of dark vibronic polaritons emerge in the system.

The presence of states $\ket{P_{ \nu \tilde\nu'}^\pm,\beta}$ could be detected by measuring the diabatic two-particle polariton splitting  
\begin{equation}\label{eq:TP splitting}
\Delta E_{\nu\tilde \nu'}=\sqrt{(N-1)}\,\Omega\,|\langle 0|\tilde \nu' \rangle|.
\end{equation}
for systems with small values of the Rabi coupling parameter $\sqrt{N}\Omega/\omega_{\rm v}\ll 1$. For nanoscale cavities having a small number of organic emitters $N<10$ within the cavity mode volume, it would be possible to measure the size-dependent ratio between the single-particle splitting $\Delta E_{\tilde \nu'}$ and two-particle polariton splitting $\Delta E_{\nu \tilde \nu'}$ associated with the {\it same} vibronic transition, which scales as
\begin{equation}
\frac{\Delta E_{\nu \tilde \nu'}}{\Delta E_{\tilde \nu'}}= 1- \frac{1}{2N}\left( 1+ \frac{1}{4N}\right).
\end{equation}
For a dimer in a cavity ($N=2$), the two-particle splitting $\Delta E_{\nu\tilde \nu'}$ is only $70\%$ the value of the single-particle splitting for any vibronic quantum number $\tilde \nu'$ and can thus be distinguished using current plasmonic cavity implementations \cite{Chikkaraddy:2016aa}. For larger samples, the single and two-particle splittings however become energetically indistinguishable. Despite this difficulty, we later envision two-color absorption experiments that can be used to detect the two-particle polariton splitting for cavities with an arbitrary number of emitters.

\subsection{Dark Vibronic Polaritons}

 As we increase the magnitude of the Rabi coupling parameter $\sqrt{N}\Omega/\omega_{\rm v}$ to values of order unity, the light-matter term of the HTC Hamiltonian [Eq (\ref{eq:HTC})] introduces couplings between diabatic single particle polaritons $\ket{P_{\tilde \nu'}^\pm}$, collective vibronic excitations $\ket{\alpha,\tilde \nu,0_c}$, and diabatic two-particle polaritons $\ket{P_{ \nu \tilde\nu'}^\pm,\beta}$ \cite{Herrera2017-PRA} that leads to the mixing of  diabatic configurations. The coupling is most efficient for values of the Rabi coupling parameter in the range $\sqrt{N}\Omega/\omega_{\rm v}\approx 1.0-2.5$ \cite{Herrera2017-PRA}, for molecular emitters with typical values of the Huang-Rhys factor $\lambda^2\approx 0.5-1.5$. %
 
\begin{figure}[t]
\includegraphics[width=0.5\textwidth]{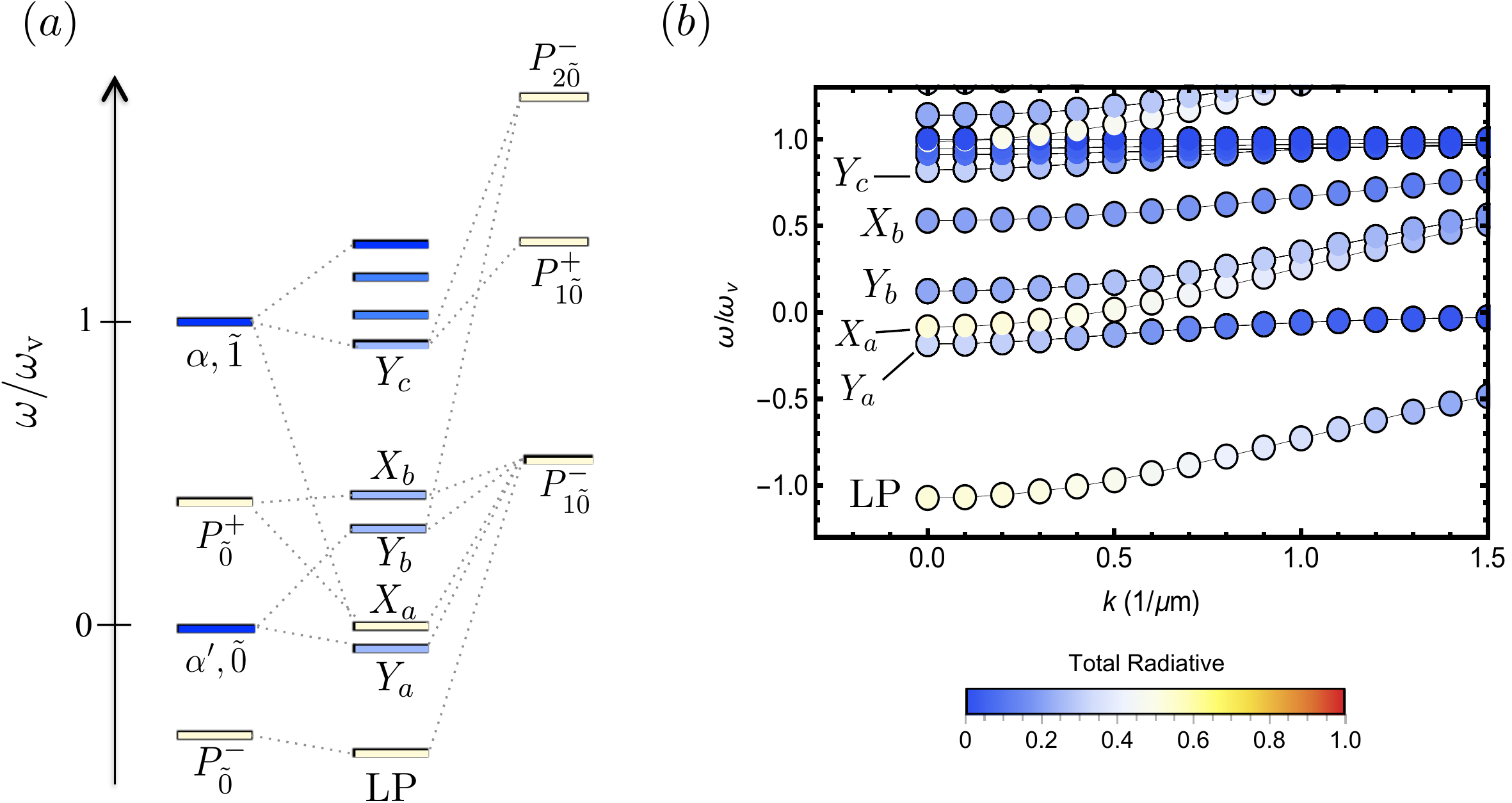}
\caption{Spectrum of the Holstein-Tavis-Cummings model for intermediate Rabi couplings $\sqrt{N}\Omega/\omega_{\rm v}\sim 1$. (a) Interaction diagram between single and two-particle diabatic states and the resulting adiabatic states. The lower polariton (LP)  and the first dark vibronic polariton states of the $X$ and $Y$ types are labelled. Only dominant light-matter couplings are explicitly shown. (b) Polariton dispersion curves for $N=10$, $\sqrt{N}\Omega/\omega_{\rm v}=2.4$ and $\lambda^2=0.5$. Symbols are colored according to the total radiative component of the corresponding polariton eigenstate. In both panels we use the resonance condition $\omega_{c}=\omega_{00}$, at normal incidence. Frequency is shown relative to $\omega_c$.}
\label{fig:HTC spectrum intermediate Rabi}
\end{figure}

 In this intermediate regime of Rabi couplings, Fig. \ref{fig:HTC spectrum intermediate Rabi} shows that the lowest four energy levels above the lower polariton $\ket{{\rm LP}}$ correspond to the dark vibronic polaritons $\ket{Y_a}$, $\ket{X_a}$, $\ket{Y_b}$, $\ket{X_b}$, whose energies are in the vicinity of bare electronic resonance frequency ($\omega_{00}=\omega_c$). In general, $X$ states are non-degenerate as they are associated with the totally-symmetric permutation quantum number $\alpha_0$ \cite{Herrera2017-PRA}. There are however $N-1$ degenerate dark vibronic polaritons of the $Y$ type associated with one of the $N-1$ possible values of the non-symmetric permutation quantum number $\alpha'\neq \alpha_0$.
 
The dispersion diagram in Figure \ref{fig:HTC spectrum intermediate Rabi}(b) shows that in the vicinity of the bare electronic (excitonic) resonance, the HTC Hamiltonian predicts the existence of large density of dark vibronic polariton states given by $\ket{Y_a}$ and $\ket{Y_b}$. The photonic part of these near-resonance $Y$ states have a significant contribution of the phonon-photon dressed state $\ket{\beta\neq \alpha_0,\nu=1,1_c}$ \cite{Herrera2017-PRA}, and their material part is mostly given by collective dark vibronic states $\ket{\beta'\neq \alpha_0,\tilde 0, 0_c}$ [Eq. (\ref{eq:dark exciton})] and the material part of the diabatic two-particle polariton $\ket{P_{1\tilde 0}^-}$ [Eq. (\ref{eq:TP polariton})]. In other words, the HTC model predicts that the so-called {\it dark exciton reservoir}, predicted to exist in this regime of Rabi couplings \cite{Agranovich2005,Litinskaya2004,Litinskaya2006}, can have a significant vibrationally-excited photonic component,  even for a homogeneous ensemble without energetic or Rabi disorder.   
 
We have previously shown that under the resonance condition $\omega_c=\omega_{00}$, it is possible to find a critical  Rabi frequency $\sqrt{N}\Omega_X$ at which the dark vibronic polariton state $\ket{X_a}$  has exactly the same energy as the bare electronic frequency $\omega_{00}$. In the rotating frame of the cavity field \cite{Herrera2017-PRA,Herrera2017-PRL}, state $\ket{X_a}$  satisfies the eigenvalue equation $\hat{\mathcal{H}}\ket{X_a}=0$, where $\hat{\mathcal{H}}$ is the HTC Hamiltonian in Eq. (\ref{eq:HTC}). This zero-energy dark vibronic polariton is interesting because it has the same energy as the uncoupled electronic transition, but nevertheless has a large {\it vibrationless} photonic component. For large particle numbers $N$, the critical coupling tends asymptotically to $\sqrt{N}\Omega_X\approx 2.4 \,\omega_{\rm v}$. For intermediate Rabi frequencies other than the critical value, the vibronic polariton state $\ket{X_a}$ remains very close to the bare electronic resonance, and the frequency of the vibronic polariton state $\ket{X_b}$ can be found in the range $\omega\approx 0.3-0.5 \,\omega_{\rm v}$, in the large $N$ limit \cite{Herrera2017-PRA,Herrera2017-PRL}. 
 
Dark vibronic polaritons of the $X$ and $Y$ types have photonic components involving vibrationally-excited photon dressed states of the form $\ket{g_m\nu_m\geq 1}\otimes \ket{1_c}$, where $m$ is a molecule index.
Due to their significant photonic component, as Fig. \ref{fig:HTC spectrum intermediate Rabi}(b) shows, $X$ and $Y$ vibronic polaritons can decay radiatively via cavity photon leakage by transmission into the far field, projecting the material into a vibrationally {\it excited} state in the ground electronic potential. The emitted photon can thus be detected at an energy $\hbar \omega $ that is {\it lower} than the energy of its parent vibronic polariton. More generally, if we denote an eigenstate of the HTC Hamiltonian within the single-excitation manifold by $\ket{\epsilon_j}$, and its eigenvalue by $\hbar \omega_j$, then we can represent a dissipative photon leakage process by the mapping

\begin{equation}\label{eq:SP loss}
\hat a \ket{\epsilon_j}\rightarrow \ket{g_1\nu_1,\ldots,g_m\nu_m,\ldots,\ldots,g_N\nu_N}\otimes \ket{0_c}+\hbar \omega,
\end{equation}
 where the energy of the emitted photon is given by
 \begin{equation}\label{eq:PL frequency}
 \hbar\omega = \hbar\omega_j - \hbar \omega_{\rm v}\sum_m\nu_m.
 \end{equation}
The single-particle approximation discussed in the previous section only allows radiative decay processes in which all the molecules are projected in their vibrationless ground state by photon leakage, i.e., $\nu_m=0$ for all $m$. However, by going beyond the single-particle approach, it is possible to treat more general emission processes in which photons detected at a given frequency $\omega$, can arise due to the decay of a vibronic polariton eigenstate at a higher energy, when $\sum_m\nu_m\geq 1$ . This general behaviour is consistent with recent measurements of the action spectra for a variety of organic emitters in metal microcavities \cite{George2015-farad,Ebbesen2016}, where emission at the lower polariton frequency grows as the cavity is pumped by monochromatic light above the lower polariton energy, in a manner that cannot be explained by non-radiative relaxation alone.

\subsection{Polaron Decoupling of the Lower Polariton Manifold}

We have shown in previous sections that for small and intermediate Rabi couplings the lower polariton state is mainly given by the single-particle state 
\begin{eqnarray}\label{eq:LP weak coupling}
\ket{P_{\tilde 0}^-} &=&\left\{\left(({1}/{\sqrt{N}})\sum_{n=1}^N\ket{g_10_1,\ldots e_n\tilde \nu_n, \ldots g_N 0_N}\right)\otimes \ket{0_c}\right.\nonumber\\
&&\left.-\ket{g_10_1,g_20_2\ldots,g_N0_N}\otimes \ket{1_c}\right\}/\sqrt{2}, 
\end{eqnarray}
which is a entangled state involving electronic, nuclear, and photonic degrees of freedom. Light-matter entanglement is a well-known feature that emerges in cavity quantum electrodynamics with atomic systems \cite{Garraway2011,Carmichael-book2}. Optical cavities with trapped atoms or ions introduce an additional vibrational degree of freedom for each particle associated with a harmonic trapping potential \cite{Beige2005,Retzker2007,RMiller2005}, which resembles the vibrational motion of the nuclei in molecules, but where the trapping potential is independent of the electronic state of the particle.  In molecular systems with vibronic coupling (see Fig. \ref{fig:displaced oscillator}), the harmonic nuclear potential is in general different for different electronic states, with different nuclear potential minima and vibrational frequencies \cite{May-Kuhn}.

\begin{figure}[t]
\includegraphics[width=0.5\textwidth]{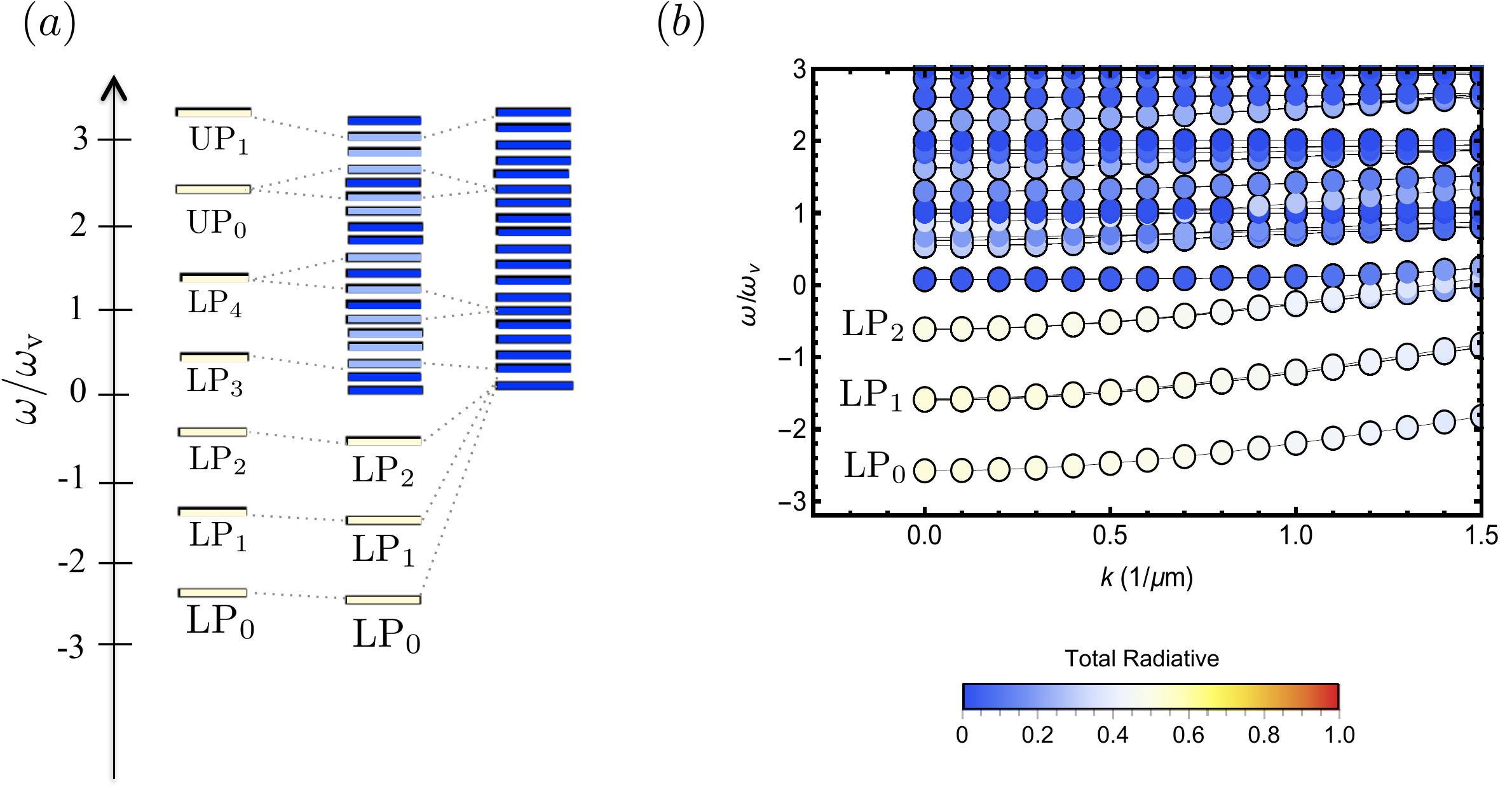}
\caption{Spectrum of the Holstein-Tavis-Cummings model in the polaron decoupling regime $\sqrt{N}\Omega/\lambda^2\omega_{\rm v}\gg 1$. (a) Interaction diagram between diabatic totally-symmetric (left side) and non-symmetric states (right side), with respect to permutation of vibronic excitation.  The lower polariton manifold (LP) is energetically separated from the non-symmetric polaron manifold. (b) Polariton dispersion curves for $N=10$, $\sqrt{N}\Omega/\omega_{\rm v}=5.5$ and $\lambda^2=0.5$. Symbols are colored according to the total radiative component of the corresponding polariton eigenstate. We assume $\omega_{c}=\omega_{00}$ at normal incidence. Frequency is shown relative to $\omega_c$.}
\label{fig:polaron decoupling spectrum}
\end{figure}

It is well-known from the standard Tavis-Cummings model for trapped atoms in an optical cavity \cite{RMiller2005} that in the absence of electron-vibration coupling, the vibrational motion of atoms in their trapping potential is separable from the internal degrees of freedom involved in light-matter coupling. Since the HTC model in Eq. (\ref{eq:HTC}) becomes formally equivalent to the Hamiltonian describing an ensemble of trapped two-level atoms in a cavity when $\lambda=0$, there must be a regime of Rabi couplings where vibronic coupling becomes negligible in comparison with the light-matter interaction, such that at least for some vibronic polariton eigenstates, the intramolecular vibrational motion becomes separable from the strongly-coupled electron-photon state, as in the atomic case. 
We have previously shown \cite{Spano2015,Herrera2016} that this  regime corresponds to the condition $\sqrt{N}\Omega/\lambda^2\omega_{\rm v}\gg 1$.  

In this so-called {\it polaron decoupling} regime \cite{Herrera2016}, we can write the lowest energy polariton eigenstates of the HTC model in Eq. (\ref{eq:HTC}) as
\begin{equation}\label{eq:PD LP}
\ket{{\rm LP}_j} =\ket{\phi_-}\otimes \hat D_{\alpha_0}^{\dagger}(\lambda_N)\ket{\nu_{\alpha_0}}\otimes \ket{\nu_{\beta_1}}\otimes \ket{\nu_{\beta_2}}\cdots\otimes \ket{\nu_{\beta_{N-1}}},
\end{equation}
where $\ket{\nu_\beta}$ is a vibrational eigenstate for the collective vibrational mode with permutation quantum number $\beta$, and the operator $\hat D_{\alpha_0}^{\dagger}(\lambda_N)$ displaces the vibrational minima of the totally-symmetric mode $\alpha_0$ by the quantity $\lambda_N$ in nuclear configuration space, relative to the equilibrium ground state nuclear configuration. The non-symmetric modes $\beta\neq \alpha_0$ are all unshifted relative to the reference configuration.   The totally-symmetric mode in Eq. (\ref{eq:PD LP} is displaced from the reference nuclear configuration by the quantity \cite{Spano2015,Herrera2016,Zeb2016,Wu2016}
\begin{equation}\label{eq:lambda N}
\lambda_N = \frac{\lambda}{2\sqrt{N}}. 
\end{equation}

Since, according to Eq. (\ref{eq:lambda N}), the equilibrium configuration of the only displaced mode in the system becomes indistinguishable from the ground state equilibrium configuration for large $N$, there is {\it no reorganization energy} for a vertical transition between the ground electronic potential and the effective lower polariton potential that characterizes the manifold in Eq. (\ref{eq:PD LP}). As a consequence, any molecular process that depends on the relative reorganization energy between ground and excited state potentials should behave very differently in a cavity in comparison with free space. We have previously shown \cite{Herrera2016} that in the polaron decoupling regime it is possible to enhance or suppress the rate of electron transfer between a donor state strongly coupled to a cavity field, and a weakly-coupled acceptor state that changes nuclear configuration upon accepting an excess electron \cite{Grabowski2003}, by a orders of magnitude. 

Equation (\ref{eq:PD LP}) shows that for  states in the lower polariton {\it manifold},  the nuclear degrees of freedom are {\it separable} from the electron-photon state, which is now given by the standard Tavis-Cummings form \cite{Garraway2011}
\begin{eqnarray}\label{eq:TC LP}
\ket{\phi_-} &=&\frac{1}{\sqrt{2}}\left[ (1/\sqrt{N})\sum_{n=1}^N\ket{g_1,\ldots e_n, \ldots g_N }\ket{0_c}\right.\nonumber\\
&&\left. -\ket{g_1,g_2,\ldots,g_N}\ket{1_c}\right].
\end{eqnarray}
In other words, there is {\it no polaron} formation due to intramolecular vibronic coupling in this regime of Rabi couplings, even for molecular systems with large Huang-Rhys factors $\lambda^2>1$ in free space.

The diabatic states in the lower polariton manifold given by Eq. (\ref{eq:PD LP}) become an accurate representation of the lowest HTC vibronic polariton eigenstates when the number of vibrational quanta in the manifold is small compared with $\sqrt{N}\Omega/2$, as illustrated in  Fig. \ref{fig:polaron decoupling spectrum}. Higher vibrational excitations of the lower polariton manifold $\ket{{\rm LP}_j}$ can energetically overlap with a high density of diabatic vibronic polariton states and form new HTC eigenstates due to interactions induced by vibronic coupling and energetic disorder \cite{Herrera2016}. The same occurs with the states in the diabatic upper polariton manifold $\ket{{\rm UP}_j}$, which are given by a product form as in Eq. (\ref{eq:PD LP}), but replacing $\ket{\phi_-}$ from (\ref{eq:TC LP}) by the orthogonal combination $\ket{\phi_+}=(\sum_n\ket{g_1,\ldots e_n, \ldots g_N}\ket{ 0_c}/\sqrt{N}+\ket{g_1\ldots g_N}\ket{1_c})/\sqrt{2}.$ 

\subsection{Radiative Lindblad Dissipation}
Up to this point, we have only described the eigenspectrum of the HTC model Hamiltonian [Eq. (\ref{eq:HTC})]. Before we can use this understanding of the polariton eigenstates to discuss the interpretation of the spectroscopic signals of organic cavities, we must first address the dynamics of dissipative processes in organic cavities. In order to do this, we adopt a standard Markovian open quantum system approach to describe the competition between coherent and dissipative processes from quantum optics  \cite{Breuer-book,Carmichael-book2}, but formulated here in terms of the vibronic polariton eigenstates of the HTC model \cite{Herrera2017-PRA,Herrera2017-PRL}.  We explicitly consider two radiative decay channels for vibronic polariton eigenstates. The first corresponds to leakage of the radiative component of a polariton state outside the spatial region where strong light-matter coupling occurs into the far field.  This type of photon decay occurs at a rate $\kappa\sim \omega_c/Q$ and determines the empty-cavity linewidth.  While for microcavities with dielectric mirrors having quality factors in the range $Q\sim 10^4-10^6$ \cite{Mabuchi2002,Tischler2007}, the photon decay time $1/\tau$ is in the order of tens or hundreds of picoseconds \cite{Muller2000}, microcavities with metallic mirrors ($Q\sim 10$) have photon decay times are only a few tens of femtoseconds \cite{Hobson2002,Ebbesen2016}, which is {\it  faster} than the typical timescales for intramolecular vibrational relaxation (IVR) \cite{May-Kuhn}. The second radiative  decay channel that we consider is the optical fluorescence of the material component of vibronic polaritons into near-resonant optical modes of the cavity that do not participate in the formation of polaritons. For microcavities, such modes are typically bound to the nanostructure by total internal reflection (bound modes), but nevertheless can carry photons away from the spatial region where strong light-matter coupling is taking place. 
%
%
Since the material component of vibronic polaritons are always delocalized over the entire molecular ensemble within our homogeneous HTC model, the corresponding superradiant fluorescence decay rate  \cite{Spano1989a} is given by $\gamma\sim N\gamma_0$, where $\tau_0\equiv1/\gamma_0$ is the fluorescence lifetime of a single organic emitter, typically in the range $\tau_0\sim 1-10$ ns. 

By writing a Lindblad master equation in the HTC polariton eigenbasis \cite{Herrera2017-PRA,Herrera2017-PRL}, with $\ket{\epsilon_j}$ representing a polariton eigenstate in the single-excitation manifold and $\ket{\epsilon_i}$ a state in the absolute ground manifold, we obtain an expression for the decay rate for the $j$-th polariton eigenstate of the form
\begin{equation}\label{eq:Gammas}
\Gamma_{j} = \kappa\sum_i|\bra{\epsilon_i}\hat a\ket{\epsilon_j}|^2+N\gamma_0\sum_j|\bra{\epsilon_i}\hat J_-\ket{\epsilon_j}|^2, 
\end{equation}
where $ \hat J_-$ is a size-normalized collective dipole transition operator \cite{Herrera2017-PRA}. This expression generalizes the dressed-state form of the polariton decay rates in the standard Tavis-Cummings model \cite{Carmichael-book2}. The first term represents the contribution to the decay from the photon leakage transitions represented by Eq. (\ref{eq:SP loss}). The second term represents the contribution to broadening due to superfluorescence into bound modes of the nanostructure.

 
\section{Organic Cavity Spectroscopy}

In organic cavities described by the HTC Hamiltonian $\hat{\mathcal{H}}$ in Eq. (\ref{eq:HTC}), we can consider the reflection and transmission spectra of a monochromatic pump $I_p(\omega)$, given by $R(\omega)$ and $T(\omega)$, respectively, as a result of dissipative photon leakage transitions occurring at the rate $\kappa$. The absorption spectra is then defined as the normalized photon flux difference $A(\omega) = 1 - R(\omega) - T(\omega)$ \cite{Savona1999}, which vanishes for an empty cavity, but can be large in strongly coupled systems due to the polariton fluorescence into bound modes of the nanostructure at the rate $N\gamma_0$, which remove photon flux from the direction of the detectors in the far field \cite{Herrera2017-PRA}.

The photoluminescence spectra $S_{\rm PL}(\omega)$ is the  result of photon leakage transitions into the far field following excitation of the cavity system with a blue-detuned weak pump field $I_p(\omega_p)$, such that $\omega_p> \omega$. We focus below on modelling stationary emission signals in which the system is continuously being driven by the weak pump, under the assumption that at most one polariton is present in the system at all times \cite{Herrera2017-PRA}. Furthermore, in order to compute the photoluminescence spectra that we assume that all energy levels within the single-excitation manifold of the HTC model are equally populated up to the energy of the pump field $\hbar \omega_p$, defining a cut-off energy above which polaritons no longer contribute to cavity emission.  

The theoretical framework described in the previous sections can also be used to describe other absorption and emission processes that are characteristic of organic cavities. These include bound absorption \cite{Herrera2017-PRA,Herrera2017-PRL} in the near field of the nanostructure \cite{Barnes1986,Matterson2001,Torma2015},  as well as bound fluorescence \cite{Spano2015} or bound photoluminescence \cite{Herrera2017-PRA}, which probe the dipole response of polaritons. In this work, we focus on the conventional cavity absorption $A(\omega)$ and photoluminescence spectra $S_{\rm PL}(\omega)$, as a function of systems parameters such as the strength of Rabi coupling, the pump frequency, and vibrational temperature, in an effort to interpret recent experimental evidence \cite{Ebbesen2016,Cuadra2016} which cannot be not fully explained using  currently available single-particle theories of organic polaritons \cite{Agranovich2003,Agranovich2005,Litinskaya2004,Litinskaya2006,Fontanesi2009,Mazza2009}. We compute the organic cavity spectra  below from analytical expressions derived previously using the quantum regression formula \cite{Herrera2017-PRA}.

\subsection{Rabi Coupling Regimes}

Let us consider the spectral response of an ideal system with $N=10$ molecular emitters in a cavity with photon and dipole decay rates that are equal in magnitude to the vibrational frequency ($\kappa=N\gamma_0=\omega_{\rm v}$), which for typical organic cavities corresponds to homogeneous broadenings on the order of a few hundred meV \cite{Ebbesen2016}. For numerical simplicity, we use a small particle number to discuss the spectroscopic signals, as it has been shown that by truncating the polariton basis within the one-excitation manifold, the polariton spectrum of finite size ensembles with $N\sim 10$ is not qualitatively different from the thermodynamic limit \cite{Herrera2017-PRA,Zeb2016}. 

\begin{figure}[t]
\includegraphics[width=0.5\textwidth]{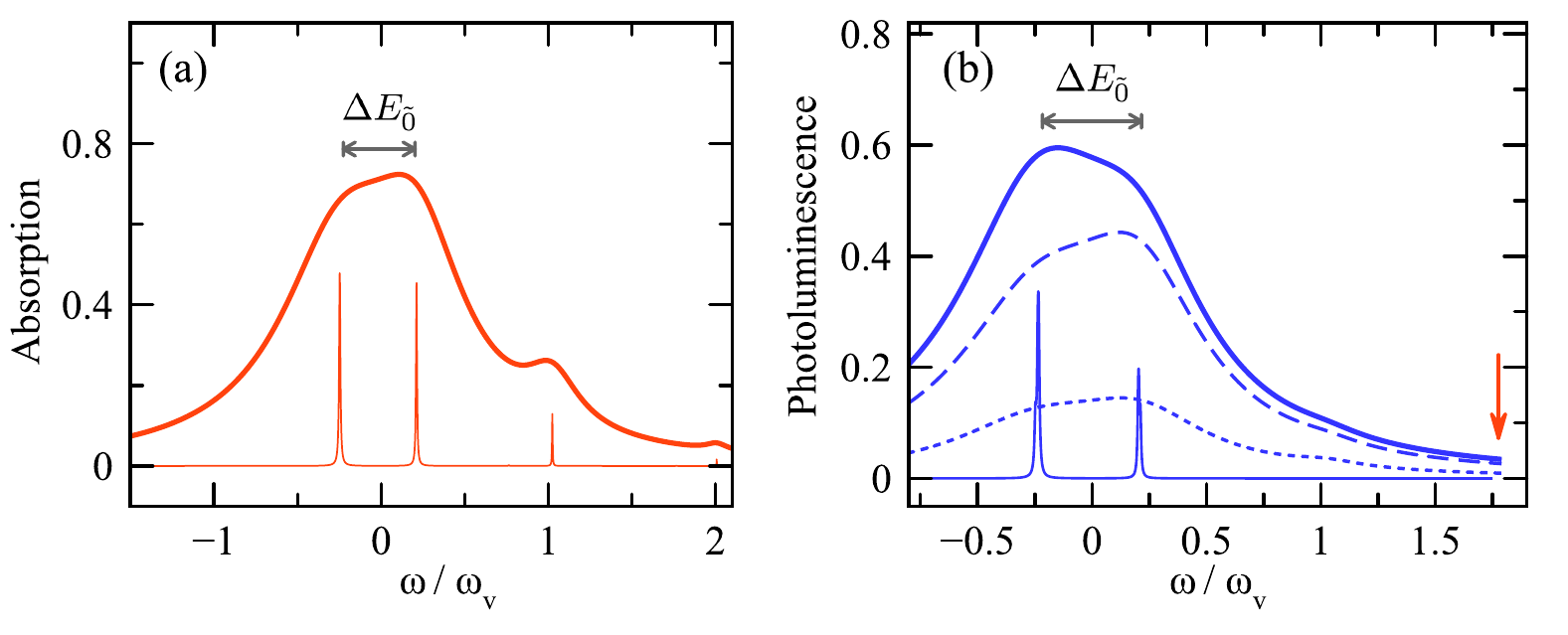}
\caption{
(a) Stationary cavity absorption $A(\omega) =1-R(\omega)-T(\omega)$ for $N$ molecular emitters representative of the small Rabi regime $\sqrt{N}\Omega/\omega_{\rm v}\ll1$, showing the diabatic single-particle splitting $\Delta E_{\tilde 0}$ between lower polariton (LP) and upper polariton states. (b) Photoluminescence spectra for the same Rabi coupling as in panel (a). Solid and dashed lines correspond to the total emission signal including photon leakage transitions that project the material into a state with up to $\nu=2$ and $\nu=1$ vibrational quanta, respectively. Dotted lines corresponds to emission events that project the system into the vibrationless ground state (single-particle approximation). The arrow indicates the cut-off frequency $\omega_{\rm cut}$ for the uniform polariton steady-state distribution.  In both panels, narrow solid lines indicate the position of the main peaks in the spectra. We use the model parameters $N=10$, $\sqrt{N}\Omega/\omega_{\rm v}=0.6$, $\kappa/\omega_{\rm v}=N\gamma_0/\omega_{\rm v}=1$, $\lambda^2=0.5$, and $\omega_c=\omega_{00}$ at normal incidence. Frequency is shown relative to $\omega_c$.
}
\label{fig:small Rabi}
\end{figure}

We begin describing the organic cavity spectra in the small Rabi limit, which we defined in previous sections by the condition $\sqrt{N}\Omega/\omega_{\rm v}<1$. Figure \ref{fig:small Rabi} shows the absorption and photoluminescence signals in the frequency region near the resonant cavity frequency $\omega_c=\omega_{00}$, which is the energy reference. All energies are given in units of the vibrational frequency $\omega_{\rm v}$.  The absorption spectra in Fig. \ref{fig:small Rabi}(a) is dominated by the conventional polariton doublet centered around the resonant frequency $(\omega=0)$, having a single-particle splitting $\Delta E_{\tilde 0}=\sqrt{N}\Omega\langle 0|\tilde 0\rangle $, which is not resolved for the chosen broadening parameters.  Figure \ref{fig:small Rabi} also shows absorption sidebands at $\omega\approx \omega_{\rm v}$ and $\omega\approx 2\omega_{\rm v}$ associated with vibronic transitions of the form $\ket{g_n0_n}\rightarrow \ket{e_n\tilde 1_n}$ and $\ket{g_n0_n}\rightarrow \ket{e_n\tilde 2_n}$, respectively. These vibronic replicas are involved in the formation of single-particle HTC polariton eigenstates for which the photonic component is dominated by the vibrationless dressed state $\hat a^\dagger\ket{G}$, as in Eq. (\ref{eq:general SP}). We note that although at the chosen value of the Rabi coupling there are diabatic two-particle polariton states $\ket{P_{1\tilde 0}^\pm}$ that are split by $\Delta E_{1\tilde 0}\approx 0.95\,\Delta E_{\tilde 0}$ around $\omega = \omega_{\rm v}$, these are not visible in  absorption at the low initial vibrational temperatures $k_{\rm b}T/\hbar \omega_{\rm v}\ll 1$, because their wavefunction does not have a contribution from the vibrationless photonic component $\hat a^\dagger\ket{G}$.  We discuss later how it would be possible to observe diabatic two-particle polariton splittings $\Delta E_{\nu \tilde\nu}$ at higher vibrational temperatures. 

The small Rabi coupling photoluminescence spectra in Fig \ref{fig:small Rabi}(b) illustrates a remarkable feature of the HTC model beyond the single-particle approximation. Since we take the cut-off frequency $\omega_{\rm cut}=1.8\omega_{\rm v}$ to be slightly more than two vibrational quanta above the lower energy peak of the absorption doublet, the emission spectrum is dominated by photons that leak out of parent polariton states $\ket{\epsilon_j}$ at higher energies $\hbar\omega_j$, which are detected in the far field at lower frequencies $ \omega=\omega_j-\sum_m\nu_m\omega_{\rm v}$, thus projecting the organic cavity into the vibrationally excited photon vacuum $\ket{g_1\nu_1,\ldots,g_m\nu_m,\ldots,\ldots,g_N\nu_N}\ket{0_c}$ with $ \sum_m\nu_m\geq  1$, as described by  Eq. (\ref{eq:SP loss}). Figure \ref{fig:small Rabi}(b) shows that within the single-particle approximation, i.e., considering only photon leakage transitions that project the system into the vibrationless ground state $\ket{G}$ in Eq. (\ref{eq:ground state}), the emission spectrum (dotted lines) reproduces the overall shape of the absorption spectrum in Fig. \ref{fig:small Rabi}(a), including the sideband structure at $\omega\approx \omega_{\rm v}$. In other words, within the single-particle approximation, a strong absorption peak would also give a strong photoluminescence peak, as expected from detailed balance. 

By going beyond the single-particle approximation, we find that even for the relatively small value of the Rabi coupling used in Fig. \ref{fig:small Rabi}, most of the photons emitted by the organic cavity at frequencies close to the bare electronic resonance are coming from parent  polariton eigenstates $\ket{\epsilon_j}$ whose energies are {\it higher} than the emitted photon. From the discussion in previous sections, those higher polaritons are mostly given by  the diabatic two-particle polariton states given by Eq. (\ref{eq:TP polariton}), in this Rabi coupling regime. 
In order to illustrate this point, Fig. \ref{fig:small Rabi}(b) shows that only 21\% of the photon flux emitted at the frequency of the lowest peak  $\omega\approx -0.23\,\omega_{\rm v}$ comes from a polariton state at that frequency ($\omega_j=\omega $). The rest of the emitted photons come from HTC polariton  eigenstates that are up to two vibrational quanta higher in energy ($\omega_j>\omega$). This behaviour can be understood by noting that the peak strength in  photoluminescence at a given transition frequency $\omega =\omega_{ji}$ is proportional to $|\langle\epsilon_i| \hat a| \epsilon_j\rangle |^2$ \cite{Herrera2017-PRA,Herrera2017-PRL}, where $\omega_{ji}$ is the frequency of the photon leakage transition involving the HTC eigenstates $\ket{\epsilon_i}$ and $\ket{\epsilon_j}$ in the absolute ground and  one-excitation manifolds, respectively. For example, the vibronic polariton states $\ket{P_{1\tilde 0}^-}$ and $\ket{P_{2\tilde 0}^-}$ can both contribute to the emission spectrum at {\it exactly} the lower polariton frequency for large $N$. The transition $\hat a\ket{P_{1\tilde 0}^-}\rightarrow \ket{\epsilon_1}$ projects the material into a state with one vibrational excitation in the ground manifold $\ket{\epsilon_1}$, and the transition $\hat a\ket{P_{2\tilde 0}^-}\rightarrow \ket{\epsilon_2}$ projects the system into a state with two vibrational excitations in the ground manifold $\ket{\epsilon_2}$. 

\begin{figure}[t]
\includegraphics[width=0.5\textwidth]{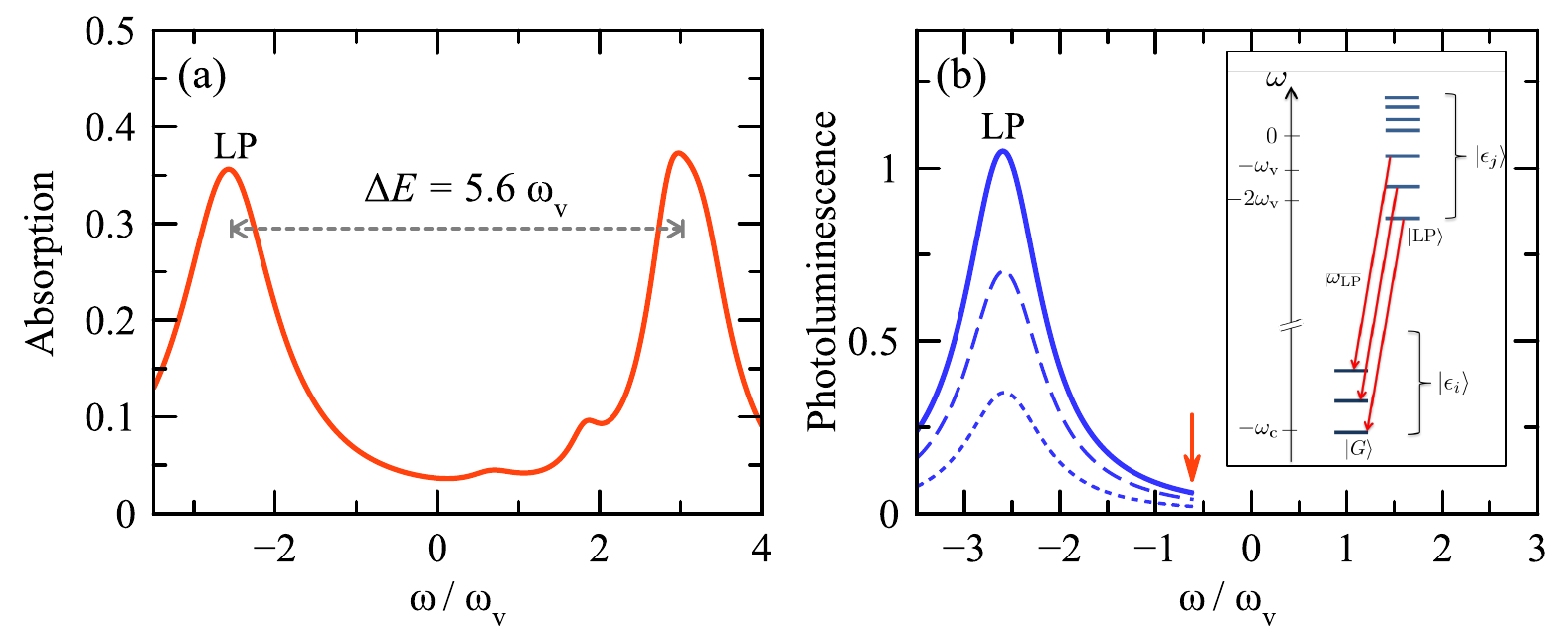}
\caption{(a) Stationary cavity absorption $A(\omega) =1-R(\omega)-T(\omega)$ for $N$ molecular emitters in the polaron decoupling regime $\sqrt{N}\Omega/\lambda^2\omega_{\rm v}\gg 1$, showing the adiabatic splitting $\Delta E$ between the lower polariton (LP) and  upper polariton peaks. (b) Photoluminescence spectra for the same Rabi coupling as in panel (a). Solid and dashed lines correspond to the total emission signal including photon leakage transitions that project the material into a state with up to $\nu=2$ and $\nu=1$ vibrational quanta, respectively. Dotted lines corresponds to emission events that project the system into the vibrationless ground state (single-particle approximation). The arrow indicates the cut-off frequency $\omega_{\rm cut}$ for the uniform polariton steady-state distribution. The inset in panel (b) shows the corresponding energy level diagram and photon leakage transitions that contribute to photoluminescence at the lower polariton frequency $\omega_{\rm LP}$.  In both panels we use the model parameters $N=10$, $\sqrt{N}\Omega/\omega_{\rm v}=5.5$, $\kappa/\omega_{\rm v}=N\gamma_0/\omega_{\rm v}=1$, $\lambda^2=0.05$, and $\omega_c=\omega_{00}$ at normal incidence. Frequency is shown relative to $\omega_c$.
}
\label{fig:large Rabi}
\end{figure}

Let us now consider the spectra of organic cavities in the opposite limit where the Rabi couplings is large, i.e., $\sqrt{N}\Omega/\omega_{\rm v}> 1$. In this regime we expect the lower polariton manifold to reach the polaron decoupled form given by Eq. (\ref{eq:PD LP}). The absorption spectrum in Fig. \ref{fig:large Rabi}(a) is indeed strongly dominated by a conventional lower and upper polariton splitting that is only slightly higher than the Tavis-Cummings value $\sqrt{N}\Omega$, confirming polaron decoupling behaviour. Deviations from ideal polaron decoupling in the upper polariton manifold result in the appearance of intermediate absorption peaks in the frequency region above the bare electronic resonance ($\omega>0$).

Figure \ref{fig:large Rabi}(b) shows the photoluminescence spectrum of the system near the lower polariton frequency, for a cut-off frequency that is two vibrational quanta above the lower polariton peak. Similar to the discussion of the small Rabi limit (Fig. \ref{fig:small Rabi}), only a small fraction of the photons emitted at the lower polariton frequency come from the lower polariton state $\ket{\rm LP}$, at that frequency. The majority of photons come from HTC polariton eigenstates at higher energies, up to the frequency cut-off frequency $\omega_{\rm cut}$ imposed by the driving field. Assuming that ideal polaron decoupling is established for the lower polariton manifold, we can use the separable wavefunction ansatz in Eq. (\ref{eq:PD LP}) to estimate the fraction of photons coming from states that are one and two vibrational quanta above the lower polariton.  According to the separable wavefunction Eq. (\ref{eq:PD LP}), the lower polariton state thus emits photons at its own energy with a strength proportional to \cite{Spano2015,Herrera2016}
\begin{eqnarray}\label{eq:PLzero strength}
|\bra{ 0_{\alpha_0}}\hat D^\dagger (\lambda_{N})\ket{0_{\alpha_0}} \braket{0_{\beta_1}|0_{\beta_1}}\cdots\braket{0_{\beta_{N-1}}|0_{\beta_{N-1}}} |^2 &&\nonumber\\
={\rm e}^{-\lambda^2/4N}&&
\end{eqnarray}
where we have used $\braket{0_{\beta_1}|0_{\beta_1}}=\cdots=\braket{0_{\beta_{N-1}}|0_{\beta_{N-1}}}=1$, and evaluated the Franck-Condon factor for the symmetric mode $\alpha_0$ using $\lambda_N=\lambda/2\sqrt{N}$, which becomes $\lambda_N\approx 0.08$ for the parameters in Fig. \ref{fig:large Rabi}.

\begin{figure}[h]
\includegraphics[width=0.4\textwidth]{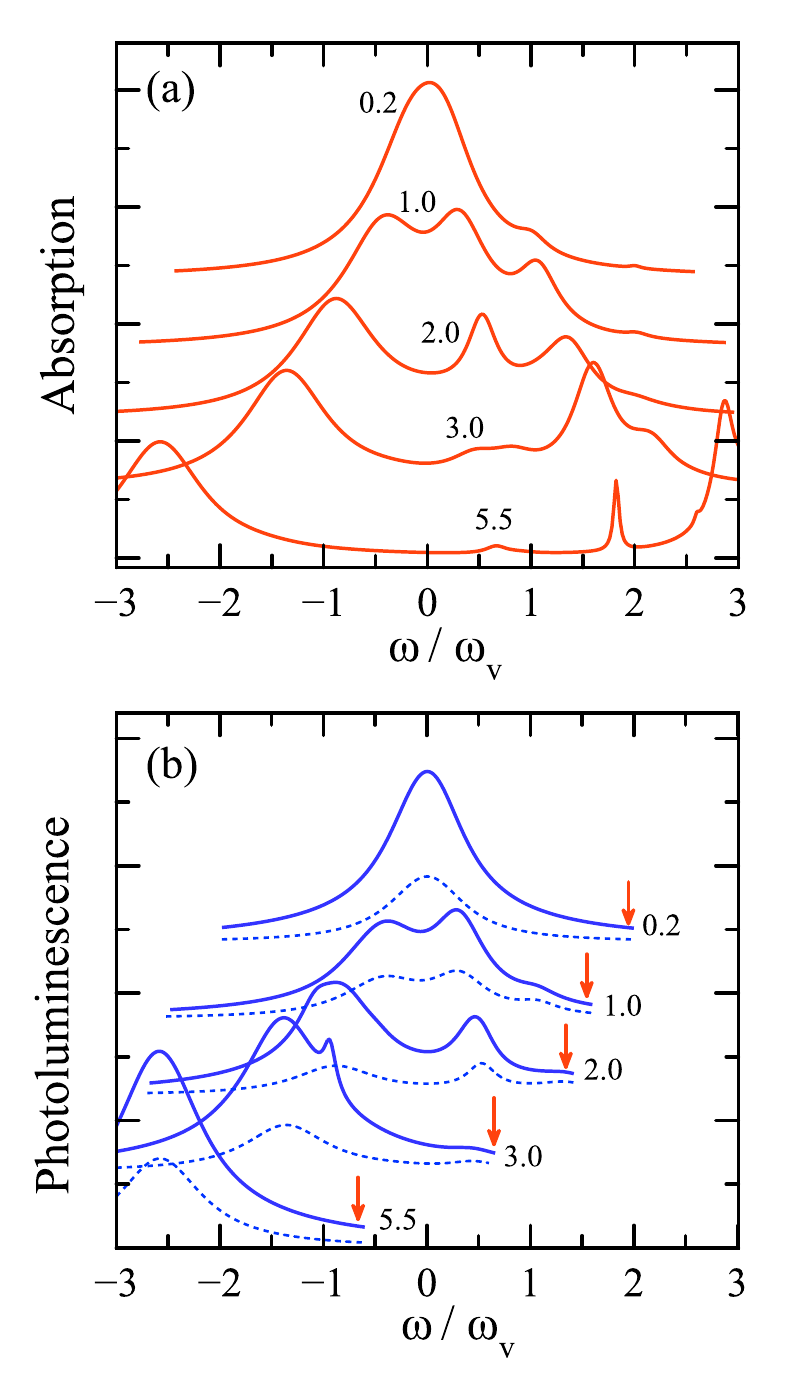}
\caption{(a) Stationary absorption spectra $A(\omega)=1-R(\omega)-T(\omega)$ for $N$ molecular emitters in a resonant cavity. Curves are labelled according to the value of the coupling ratio $\sqrt{N}\Omega/\omega_{\rm v}$.  (b)  Photoluminescence spectra for the same coupling ratios as in Panel (a).  Solid lines correspond to the total emission signal including photon leakage transitions that project the material into a state with up to $\nu=2$ vibrational quanta. Dotted lines corresponds to emission events that project the system into the vibrationless ground state (single-particle approximation). Arrows indicate the high energy cut-off  $\hbar\omega_{\rm cut}$ of the uniform polariton steady-state distribution.For $\sqrt{N}\Omega/\omega_{\rm v}=2$, there is a vibronic polariton state at $\omega\approx 0.1$ that is completely dark in absorption, but contributes to emission near the lower polariton frequency. In both panels we use the parameters
$N=10$, $\kappa/\omega_{\rm v}=N\gamma_0/\omega_{\rm v}=1$, $\lambda^2=0.5$, and $\omega_c=\omega_{00}$ at normal incidence.}
\label{fig:varying Rabi}
\end{figure}

Leakage emission from a state in the lower polariton manifold that is exactly one vibrational quanta above the lower polariton is proportional to the square of the vibrational overlap $\bra{ 0_{\alpha_0}}\hat D^\dagger (\lambda_{N})\ket{0_{\alpha_0}} \braket{1_{\beta_1}|1_{\beta_1}}\cdots\braket{0_{\beta_{N-1}}|0_{\beta_{N-1}}}$, which has the same value as in Eq. (\ref{eq:PLzero strength}).
We have assumed that mode $\beta_1$ is excited both in the parent polariton state $\ket{\epsilon_j}$ and the ground manifold state $\ket{\epsilon_i}$, but any choice of $\beta$ satisfying the constraint $\sum_\beta\nu_\beta=1$ gives the same result. If the vibrational excitation is assumed to be in the totally-symmetric mode, the overlap becomes $\bra{ 1_{\alpha_0}}\hat D^\dagger (\lambda_{N})\ket{1_{\alpha_0}}={\rm e}^{-\lambda^2/4N}(1-\lambda^2/4N)$, which tends to Eq. (\ref{eq:PLzero strength}) for large $N$. The inset in Figure \ref{fig:large Rabi}(b) illustrates that for polaritons whose photonic component has one quantum of vibration, photon leakage leads to emission detected also at the lower polariton frequency.  Similarly, leakage emission from the polariton state $\ket{\epsilon_j}$ that is exactly two vibrational quanta above the lower polariton, emits radiation at the lower polariton frequency with a strength that is proportional to the square of a vibrational overlap that has the same value as in Eq. (\ref{eq:PLzero strength}). In summary, only one third of the emitted photons at the lower polariton frequency come from directly from the lower polariton state in the polaron decoupling regime, when the system in pumped two vibrational quanta above the lower polariton. The remaining two thirds come in equal contributions from the first and second excited sideband of the lower polariton manifold given by Eq. (\ref{eq:PD LP}), behaviour that is clearly captured in Fig. \ref{fig:large Rabi}(b).


In the intermediate Rabi coupling regime (see Fig. \ref{fig:HTC spectrum intermediate Rabi}), stationary cavity absorption close to the resonance frequency ($\omega\approx 0.4\,\omega_{\rm v}$ for $\sqrt{N}\Omega=2\,\omega_{\rm v}$) is largely due to the vibrationless photonic component of state $\ket{X_b}$ in Fig. \ref{fig:HTC spectrum intermediate Rabi}, which adiabatically connects with the single-particle diabatic upper polariton state $\ket{P_{\tilde 0}^+}$ from Eq. (\ref{eq:SP polariton}), as $\sqrt{N}\Omega$ decreases. When the ratio $\sqrt{N}\Omega/\omega_{\rm v}$ exceeds unity and the lower polariton manifold begins to enter the polaron decoupling regime, the vibronic polariton state $\ket{X_b}$ acquires a stronger two-particle polariton character involving vibrationally-excited photonic components, making it more difficult to populate directly in a one-photon transition starting from the absolute ground state $\ket{G}$. The mid-region absorption peak thus tends to disappear for larger values of the Rabi coupling. For example, $X$-state absorption at $\omega=0.6\,\omega_{\rm v}$ in Fig. \ref{fig:varying Rabi}(a)  is an order of magnitude weaker than the lower polariton absorption at $\omega =-2.6\,\omega_{\rm v}$ for $\sqrt{N}\Omega=5.5\,\omega_{\rm v}$.

As discussed earlier,  the lineshape of the  stationary photoluminescence spectra strongly depends on the cut-off energy $\hbar\omega_{\rm cut}$ over which no polariton eigenstate is significantly populated in the steady state.  In Fig. \ref{fig:varying Rabi}(b), we show the static photoluminescence spectra assuming a {\it uniform population} \cite{Herrera2017-PRA} of polariton energy levels up to a cut-off frequency $\omega_{\rm cut}$ that is roughly two vibrational quanta above the lower polariton frequency, for the same values of the Rabi coupling ratio $\sqrt{N}\Omega/\omega_{\rm v}$ as in Fig. \ref{fig:varying Rabi}(a). The dominant contribution to photoluminescence at the frequency of the lowest energy absorption peak (lower polariton) comes from the {\it radiative} decay of HTC polariton eigenstates that are one or two vibrational quanta higher in energy, via photon leakage  transitions represented by Eq. (\ref{eq:SP loss}). 

\subsection{Hot-Band Cavity Absorption}

We have previously derived an expression for the absorption spectrum $A(\omega) = 1-R(\omega)-T(\omega)$ at zero temperature  \cite{Herrera2017-PRA,Herrera2017-PRL}, by considering the perturbative action of an external periodic driving term given by $\hat V_p(t)=\Omega_p(\hat a^\dagger\,{\rm e}^{-i\omega t}+\hat a\,{\rm e}^{i\omega t})$ on the HTC Hamiltonian [Eq. (\ref{eq:HTC})], together with the quantum regression formula under weak driving conditions $\Omega_p\ll \sqrt{N}\Omega$ \cite{Carmichael-book1}. At zero vibrational temperature, i.e., $k_{\rm b}T/\hbar\omega_{\rm v}\ll 1$ we expect the cavity to be in the absolute ground state $\ket{G}\equiv \ket{g_10_1,g_20_2\ldots,g_N0_N}\ket{0_c}$ in the absence of the external field. Continuous resonant driving with an optical field at frequency $\omega$ can thus induce transitions into polariton eigenstates  $\ket{\epsilon_j}$, which we represent by the mapping $\hat V_p\ket{G}\rightarrow \ket{\epsilon_j}$. In order to contribute to the absorption signal, the weakly-populated polariton state $\ket{\epsilon_j}$ can then decay by fluorescence \cite{Spano2015} into a near-resonant bound modes of the nanostructure \cite{Barnes1986,Barnes1998,Matterson2001}, thus reducing the photon flux direction where reflection $R(\omega)$ and transmission $T(\omega)$ is collected in the far field \cite{Carmichael-book2}, assuming a near-unity polariton quantum yield.

\begin{figure}[t]
\includegraphics[width=0.5\textwidth]{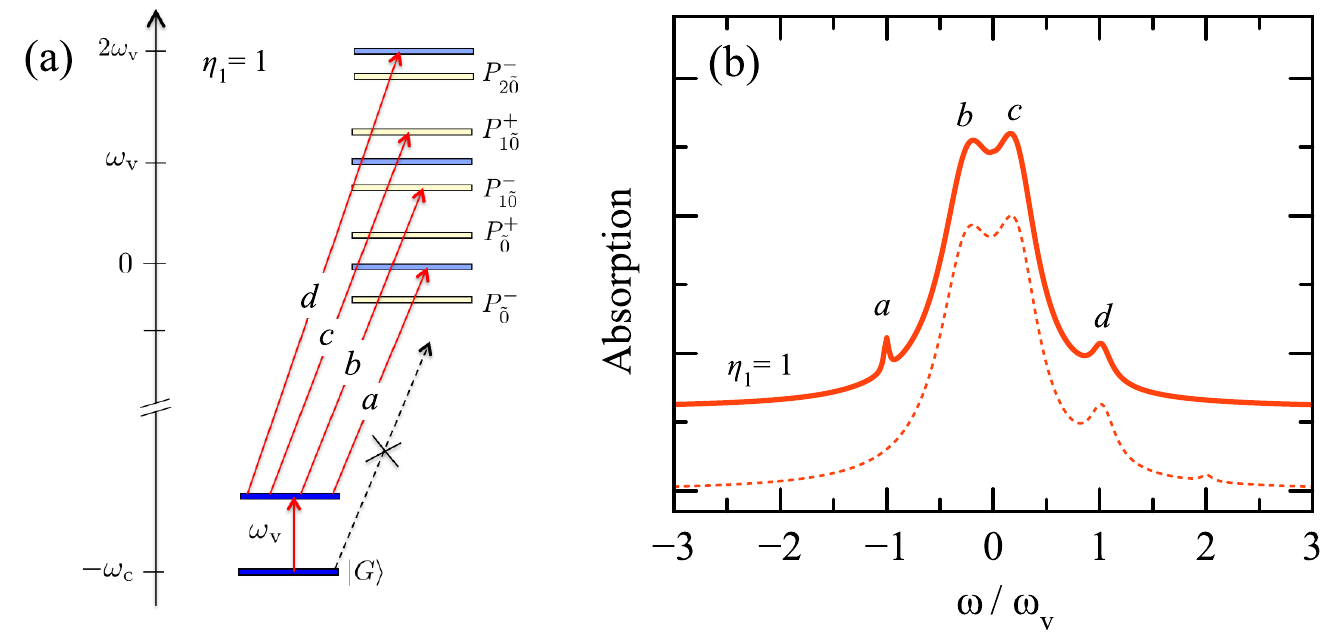}
\caption{(a) Two-color scheme to measure hot-band cavity absorption, indicating the allowed optical transitions in a system where the population is put initially in the first vibrational level ($\eta_1=1$) by an infrared field at frequency $\omega=\omega_{\rm v}$ that removes population from the absolute ground state of the cavity $\ket{G}$. For the chosen system parameters, there are four allowed transitions that result in cavity absorption, labelled {\it a} to {\it d}. Transitions from the absolute ground state $\ket{G}$ below the lower polaron state $\ket{P_{\tilde 0}^-}$ are not possible (dashed arrow). (b) Cavity absorption spectrum for a system initially in the first excited level of the ground manifold (solid line), and absorption from the absolute ground state (dashed line). Absorption peaks are labelled according to the diagram in panel (a). We use the parameters
$N=10$, $\sqrt{N}\Omega/\omega_{\rm v}=1.0$, $\kappa/\omega_{\rm v}=N\gamma_0/\omega_{\rm v}=0.7$, $\lambda^2=0.5$, and $\omega_c=\omega_{00}$ at normal incidence.}
\label{fig:pump probe}
\end{figure}

In general, the cavity system in the absence of external driving can be described by the slightly more general state
\begin{equation}\label{eq:rho mixed}
\hat \rho_0 = \sum_i\eta_i\ket{\epsilon_i}\bra{\epsilon_i},
\end{equation}
where the real quantities $\eta_i$ determine the probability distribution of the ground manifold states $\ket{\epsilon_i}$ in the ensemble. If we denote by $\hbar\omega_i$ the energy of state $\ket{\epsilon_i}$ in the ground manifold, and  $\hbar\omega_j$ the energy of the polariton eigenstate $\ket{\epsilon_j}$, then we can use Eq. (\ref{eq:rho mixed}) to extend our zero-temperature results \cite{Herrera2017-PRA,Herrera2017-PRL}, and write the stationary absorption signal as
 \begin{equation} \label{eq:absorption}
 A(\omega) =\pi\sum_i\eta_i\sum_jF_j\frac{|\langle \epsilon_i|\hat a|\epsilon_j\rangle|^2 (\kappa_{ij}/\Gamma_j)}{(\omega-\omega_{ji})^2+\kappa_{ij}^2},
 \end{equation}
where $\kappa_{ij}$ is the decay rate for the coherence between states $\ket{\epsilon_i}$ and $\ket{\epsilon_j}$, $\Gamma_j$ is the polariton radiative decay rate given by Eq. (\ref{eq:Gammas}), and $\omega_{ji}=\omega_j-\omega_i$ is the transition frequency. The total dipole emission strength of the $j$-th polariton eigenstate, $F_j = \sum_i|\langle\epsilon_i|\hat \mu| \epsilon_j\rangle|^2$, accounts for the fact that polaritons which cannot fluoresce into bound modes do not contribute to the far-field absorption spectra.

The expression for the stationary cavity absorption in Eq. (\ref{eq:absorption}) can be used to interpret experiments where the vibrational temperature of the relevant vibrational mode is not negligible, i.e., $k_{\rm b}T/\hbar\omega_{\rm v}\sim 1$, by associating the population factors $\eta_i$ with a properly normalized statistical distribution. The mixed state $\hat \rho_0$ in Eq. (\ref{eq:absorption}) can also describe a non-thermal state corresponding to a cavity under strong infrared driving that is resonant with an excited vibrational level at frequency $\omega=\nu\omega_{\rm v}$, hosting $\nu$ vibrational quanta in the ensemble. Under such conditions, the population of the absolute ground state can become negligible over the relevant driving timescales, i.e., $\eta_G\ll 1$, such that the absorption of a second driving field at optical frequencies $\omega\sim \omega_{ji}$ is dominated by transitions starting from an excited vibrational state in the ground manifold, for which $\eta_i \sim 1$, leading to a type of hot-band cavity absorption. A continuous-wave version of the envisioned IR-optical two-photon cavity absorption sequence is illustrated in Fig. \ref{fig:pump probe}(a). In experiments, pulsed infrared and optical fields could be considered, and the initial state in Eq. (\ref{eq:rho mixed}) could be generalized to account for laser-induced vibrational coherences in the ground manifold. 

Figure \ref{fig:pump probe} shows that for relatively small Rabi couplings $\sqrt{N}\Omega<\omega_{\rm v}$, optical cavity absorption starting from the first excited vibrational level of the ground manifold can probe directly the two-particle polariton splitting $\Delta E_{\nu \tilde \nu'}$, given by Eq. (\ref{eq:TP splitting}), provided the splitting can be resolved. In experiments where single-particle polariton splittings have been resolved for a small number of organic emitters $N\sim 1-10$ \cite{Chikkaraddy:2016aa}, it should thus be possible to measure the anharmonicity of the light-matter coupling characterized by the $\sqrt{(N-1)}$ scaling of the two-particle polariton splitting $\Delta E_{\nu \tilde \nu'}$, provided that the hot-band contribution to the absorption spectra can be isolated, as in the two-color scheme proposed here. 

More generally, hot-band absorption could  be used to spectroscopically detect the presence of any HTC polariton eigenstate with negligible {\it vibrationless} photonic component, such as dark-vibronic polaritons of the $Y$ type at intermediate Rabi couplings ($\sqrt{N}\Omega/\omega_{\rm v}\sim1$), as well as excited vibrational sidebands of the lower polariton manifold $\ket{{\rm LP}_{j\geq 1}}$ [Eq. (\ref{eq:PD LP})] in the polaron decoupling regime  ($\sqrt{N}\Omega/\lambda^2\omega_{\rm v}\gg1$). Figure \ref{fig:pump probe}(b) also shows that hot-band absorption starting from the first vibrationally excited state in the ground manifold gives a sideband absorption peak in the region  {\it below} the conventional lower polariton peak. For the system parameters chosen in Fig. \ref{fig:pump probe}, this lower energy sideband is exactly one vibrational quanta below the bare resonance frequency $\omega_c=\omega_{00}$, and its strength is proportional to the vibrationally excited photonic component of the non-symmetric collective vibronic states $\ket{\alpha',\tilde 0, 0_c}$, the so-called dark exciton states [see Eq. (\ref{eq:dark exciton})], at the bare electronic resonance. Fig. \ref{fig:pump probe} also shows that there should be no sideband structure below the lower polariton frequency when absorption is {\it exclusively} due to  transition from the absolute ground state $\ket{G}$. 


\section{Conclusion and Outlook}

In this work, we have described the general photophysics of organic cavities using a homogeneous Hostein-Tavis-Cummings (HTC) model \cite{Cwik2014,Spano2015,Herrera2016,Herrera2017-PRA,Herrera2017-PRL} [Eq. (\ref{eq:HTC})] to describe the structure of vibronic polaritons, together with a Markovian description of their radiative decay dynamics. The developed theoretical framework provides a microscopic interpretation for some of the  observed features in the absorption and emission spectra of organic microcavities over a range of light-matter coupling strengths within the so-called strong coupling regime of cavity quantum electrodynamics. 
The model generalizes previous theoretical approaches that described spectroscopic signals under the assumption that the photonic part of the polariton wavefunction has no vibrational structure, but only the material part of the wavefunction can have the contribution of several vibronic transitions \cite{Agranovich2003,Agranovich2005,Litinskaya2004,Litinskaya2006,Fontanesi2009,Mazza2009}. We refer to this as the single-particle approach to the description of vibronic polaritons. 

Within this single-particle approach, photoluminescence at the lower polariton frequency $\omega_{\rm LP}$ can only be due to the radiative decay of the lower polariton state $\ket{\rm LP}$ via leakage of the photonic component of its wave function into the far field. In this picture, when a cavity is driven with a narrow laser field above the lower polariton frequency, the higher-energy states that are populated by the external driving must first decay via phonon-induced non-radiative transitions into the lower polariton state, before radiative decay can take place \cite{Litinskaya2004,Litinskaya2006,Fontanesi2009,Mazza2009}. As the system relaxes towards the lower polariton state, it is assumed that non-radiative relaxation populates a large number of weakly-coupled collective material states, the so-called excitonic reservoir \cite{Litinskaya2004,Litinskaya2006}, in the frequency region near the bare electronic (excitonic) resonance. Such excitonic reservoir states cannot form polaritons because of quasi-momentum mismatch, but can nevertheless act as a scattering reservoir for polariton states. In particular, it is has been proposed \cite{Fontanesi2009,Mazza2009} that material states in the excitonic reservoir can radiatively decay via cavity-enhanced fluorescence (Purcell effect) into the first excited vibration in the ground electronic molecular potential, such that the emitted photon can be reabsorbed by the ensemble and  incoherently populate any polariton state that is found one quantum of vibrational energy lower than fluorescent exciton reservoir state. The lower energy polariton state populated by such exciton-mediated scattering process can then decay radiatively via photon leakage into the far field and contribute to the photoluminescence signal. This phenomenology has been used to interpret the experimental observation of enhanced photoluminescence at the lower polariton frequency $\omega_{\rm LP}$ in cavities where the bare electronic (excitonic) resonance is roughly one vibrational quantum above the lower polariton state \cite{Coles2011,Virgili2011}. In our language, this condition is fulfilled at intermediate Rabi coupling parameters $\sqrt{N}\Omega/\omega_{\rm v}\approx 2$, for typical strengths of vibronic coupling.

Throughout this work we have shown that the HTC model of organic cavities provides a profoundly different view of polariton emission from the previously proposed interpretations. If we consider emission at the lower polariton frequency $\omega_{\rm LP}$, our theory shows that by  continuously driving a cavity with a pump energy at least one vibrational quantum above the lower polariton state, only {\it a minority of the emitted photons} at $\omega_{\rm LP}$ originate directly from the radiative decay of the lower polariton state. Most of the emitted photons at $\omega_{\rm LP}$  originate from the {\it direct radiative decay} of vibronic polariton eigenstates at {\it higher} energies. Most importantly, we demonstrate in this Review that such emission behaviour is {\it universal} in the sense that it occurs for any value of the Rabi frequency, as long as the cavity can transiently support the formation of vibrationally-excited single-photon dressed states. This proposed view of radiative polariton decay may have important implications in our understanding of macroscopic coherence phenomena such as condensation \cite{Deng2010,Daskalakis2017} and lasing \cite{Kena-Cohen2010,Ramezani2017}.

In an organic cavity where the Rabi frequency $\sqrt{N}\Omega$ is not much greater than the typical vibrational relaxation rates \cite{Lidzey2000,Holmes2007,Tischler2007}, a vibrationally-excited photonic dressed state can relax non-radiatively before it can exchange energy with purely material excitations through coherent light-matter Rabi  coupling. Under such conditions, two-particle polaritons as described in this review cannot be sustained, and our theory  of organic cavities reduces to the more commonly used single-particle approach \cite{Agranovich1997,Agranovich2003,Litinskaya2004,Litinskaya2006,Cwik2016,Fontanesi2009,Mazza2009}. The proposed theoretical framework based on two-particle and multi-particle vibronic polariton states, theory thus relies on  the ability of vibration-photon dressed states to resonantly couple with material vibronic-vibrational excitations over sub-picosecond timescales, which is the typical magnitude of the non-radiative relaxation time $\tau_{\rm nr}$ in organic systems \cite{May-Kuhn}. In general, if $\kappa$ is the decay rate for cavity photons due to  leakage into the far field, the new type of dark vibronic polaritons described in this review can be sustained in organic cavities with system parameters that satisfy the conditions $\sqrt{N}\Omega \gg \kappa\gg 1/\tau_{\rm nr}$.  These conditions hold for several experimental realizations of organic cavities with different geometries and material composition.

The qualitatively new description of organic cavities introduced by the homogeneous HTC model can be further refined by introducing more detailed features of organic cavities such spatial inhomogeneities in the electronic transition frequency \cite{Spano2015,Herrera2016}, inhomogenous Rabi coupling \cite{Gonzalez2013}, and a more elaborate treatment of the multi-mode nature of the commonly used photonic nanostructures than the quasi-mode approximation used here. Since the typical radiative and non-radiative molecular relaxation channels are well-known, it should be also possible to develop a microscopic dynamical description of the competition between Rabi coupling, cavity photon leakage, cavity-enhanced molecular fluorescence and vibrational relaxation that can go beyond the steady-state description used in this work. Such dynamical open quantum system modelling would allow us to better understand ultrafast cavity spectroscopy measurements \cite{Ebbesen2016}, as well as the temperature dependence of organic polariton signals \cite{Cuadra2016}.

By further exploring the condensed-matter aspects of the HTC model, it should also be possible to describe organic cavities under strong resonant and non-resonant driving, in order to better understand the quantum non-linear dynamical processes that emerge at large polariton densities, such as condensation \cite{Deng2010,Daskalakis2017} and lasing \cite{Kena-Cohen2010,Ramezani2017}. It was in this context that the HTC model was first introduced \cite{Cwik2014}, but still several open questions remain regarding the mechanisms that allow the emergence stationary macroscopic polariton coherence \cite{Lerario2017}. The cavity quantum electrodynamics framework implicit in the HTC model, also suggests possible connections with macroscopic phase coherence effects such as quantum synchronization \cite{Zhu:2015}.

In summary, the quantum theory of vibronic polaritons and organic cavity spectroscopy described in this Review, can not only serve as a phenomenological benchmark for new electronic structure calculation methods that treat the strong electron-photon coupling in confined electromagnetic fields from first principles \cite{Flick2017}, but also stimulate the development  of novel nonlinear optical devices  \cite{Herrera2014,kowalewski2016cavity,Saurabh2016,Kowalewski2016}, chemical reactors  \cite{Hutchison:2012,Herrera2016,Galego2016}, and optoelectronic devices  \cite{Feist2015,Schachenmayer2015,Orgiu2015,yuen2016} that can be enhanced by quantum optics.

\acknowledgements
F. Herrera is supported by Proyectos Basal USA 1555--VRIDEI 041731.  F. C. Spano is supported by the NSF, Grant DMR--1505437.


\bibliography{review}

\end{document}